\documentclass[conference]{IEEEtran}
\IEEEoverridecommandlockouts

\usepackage[T1]{fontenc}

\usepackage{cite}
\usepackage{amsmath,amssymb,amsfonts}
\usepackage{algorithmic}
\usepackage{graphicx}
\usepackage{textcomp}

\usepackage{makecell}

\usepackage{svg}
\usepackage{booktabs}
\usepackage{xcolor}

\usepackage{tabularx}

\usepackage{caption}
\usepackage{subcaption}

\usepackage{amssymb}
\usepackage{amsmath}
\usepackage{siunitx}

\usepackage{nicematrix}
\usepackage{tikz}

\usepackage{multicol}
\usepackage{multirow}

\usepackage{marvosym}

\usepackage{xurl}

\usepackage{seqsplit}

\usepackage{lineno}

\usepackage[hidelinks]{hyperref}

\def\NProjs/{3494}
\def\NProjsUseEth/{939}
\def\NProjsParsed/{829}
\def\NProjsErrors/{90}
\def\NProjsUseHosts/{99}
\def\sooo/{22}
\def\sooi/{26}
\def\soio/{62}
\def\sioo/{729}
\def\soii/{0}
\def\sioi/{72}
\def\siio/{27}
\def\siii/{1}

\def\BibTeX{{\rm B\kern-.05em{\sc i\kern-.025em b}\kern-.08em
    T\kern-.1667em\lower.7ex\hbox{E}\kern-.125emX}}

\makeatletter 
\newcommand{\linebreakand}{%
\end{@IEEEauthorhalign}
\hfill\mbox{}\par
\mbox{}\hfill\begin{@IEEEauthorhalign}
}
\makeatother 

\begin{document}

\title{Towards Verifiability of Total Value Locked (TVL) in Decentralized Finance}

\author{\IEEEauthorblockN{1\textsuperscript{st} Pietro Saggese}
\IEEEauthorblockA{\textit{IMT School for Advanced Studies Lucca}\\
Lucca, Italy \\
pietro.saggese@imtlucca.it
}
\and
\IEEEauthorblockN{2\textsuperscript{nd} Michael Fr{\"o}wis}
\IEEEauthorblockA{\textit{Iknaio Cryptoasset Analytics}\\
Vienna, Austria  \\
michael@ikna.io}
\and
\IEEEauthorblockN{3\textsuperscript{rd} Stefan Kitzler}
\IEEEauthorblockA{\textit{Complexity Science Hub} \\ 
Vienna, Austria \\
kitzler@csh.ac.at}
\and

\linebreakand

\IEEEauthorblockN{4\textsuperscript{th} Bernhard Haslhofer}
\IEEEauthorblockA{\textit{Complexity Science Hub} \\ 
Vienna, Austria \\
haslhofer@csh.ac.at}
\and
\IEEEauthorblockN{5\textsuperscript{th} Raphael Auer}
\IEEEauthorblockA{\textit{Bank for International Settlements} \\
Basel, Switzerland \\
raphael.auer@bisih.org}
}

\maketitle
\thispagestyle{plain}
\pagestyle{plain}


\begin{abstract}

Total Value Locked (TVL) aims to measure the aggregate value of cryptoassets deposited in Decentralized Finance (DeFi) protocols. 
Although blockchain data is public, the way TVL is computed is not well understood. In practice, its calculation on major TVL aggregators relies on self-reports from community members and lacks standardization, making it difficult to verify published figures independently. 
We thus conduct a systematic study on 939 DeFi projects deployed in Ethereum. We study the methodologies used to compute TVL, examine factors hindering verifiability, and ultimately propose standardization attempts in the field.
We find that 10.5\% of the protocols rely on external servers; 68 methods alternative to standard balance queries exist, although their use decreased over time; and 240 equal balance queries are repeated on multiple protocols. These findings indicate limits to verifiability and transparency. 
We thus introduce ``verifiable Total Value Locked'' (vTVL), a metric measuring the TVL that can be verified relying solely on on-chain data and standard balance queries.
A case study on 400 protocols shows that our estimations align with published figures for 46.5\% of protocols. Informed by these findings, we discuss design guidelines that could facilitate a more verifiable, standardized, and explainable TVL computation.

\end{abstract}



\vspace{-0.2cm}

\section{Introduction}
\label{sec:intro}

The core value proposition of Decentralized Finance (DeFi) 
lies in its transparency and reliance on permissionless on-chain infrastructure~\cite{auer2024technology}. As Total Value Locked (TVL) 
has become a primary metric for assessing the economic scale of DeFi --- measuring the value of assets deposited in smart contracts --- it is essential that its calculation upholds the same principles, remaining fully anchored in on-chain data and fully reproducible calculations.
In spite of this, TVL computation today suffers from a lack of standardization and is difficult to verify independently.

In practice, TVL estimates are published by aggregators like DappRadar, Stelareum, and DeFiLlama, which typically adopt a community-driven approach. 
The latter, for instance, allows anyone to integrate a new DeFi protocol and its TVL computation methodology by developing a protocol-specific plugin and publishing it in an open-source GitHub repository~\cite{defillama_adapters}. 
Contributors are encouraged to use only on-chain data for TVL calculations~\cite{defillama_noapis}. However, some plugins rely on data from external services and use self-defined functions for on-chain computations.
Moreover, there is variability in how the values of assets deposited in contract accounts are calculated. At the end of 2024, Ethereum TVL estimates published by different aggregators vary from approximately \$80 bln to \$190 bln, indicating that remarkable differences exist in the methodologies used for TVL computation.

The need for independent verifiability has been demonstrated in cases where DeFi developers on the Solana blockchain deliberately designed their protocols to inflate the actual value of deposited assets, ultimately manipulating TVL figures~\cite{nelson:2022a}. The necessity for a standardized approach has become evident with the recognition that crypto deposits can be double-counted across DeFi protocols~\cite{chiu2023understanding,coindesk_toggles,nuzzi:2021a}. 
This issue has recently been addressed by proposing an alternative metric, Total Value Redeemable (TVR), which refines TVL by excluding cryptoassets that derive their value from underlying cryptoassets~\cite{luo2024piercing}. However, a comprehensive understanding of the methodologies used to compute TVL is still lacking. 

In this paper, we aim to fill this gap by conducting a comprehensive measurement study to examine how TVL is computed in practice and to assess the extent to which its value can be recomputed and verified using on-chain data only. 
Our contributions and key empirical findings are as follows:

\begin{enumerate}

	\item We develop and apply a measurement instrument to $\NProjsUseEth/$ DeFi protocols on the Ethereum chain. Our findings reveal that (i) $10.5\%$ of them rely partially or entirely on data from external services to compute TVL, impeding full reproducibility; (ii) while the majority of protocols ($78.6\%$) use standard balance queries, a subset of non-standard, self-defined balance functions ($N = 68$) is employed; (iii) the usage of these alternative queries has declined from $28.2\%$ in January 2023 to $8.9\%$ in January 2024; (iv) $240$ balance queries executed on the same contracts and tokens are repeated on multiple protocols.
	
	\item We introduce \emph{verifiable Total Value Locked (vTVL)}, a metric assessing to what extent individual projects' TVL can be reconstructed using blockchain data and standard balance queries, and the \emph{Discrepancy Ratio}, to quantify differences between published data and our estimates. 
	A case study on 400 protocols shows that discrepancies are negligible for 23.5\%, and estimations align with published figures for another 23\%. 
	\item We propose design guidelines, informed by the challenges we identified, that could lead to a more verifiable and standardized TVL computation: (i) compute TVL from on-chain sources; (ii) publish protocol-specific contracts and tokens lists; (iii) favor standard balance methods whenever possible; (iv) publish the token categorizations used; and (v) define common standards for protocol selection criteria and version management.
\end{enumerate}

As DeFi continues to mature, it is essential to rely on clear and reproducible metrics, especially when these are heavily used for business and investment decisions. Standardization is crucial in this regard, and verifiability should be part of a broader discussion on auditing within the DeFi ecosystem. 

Section~\ref{sec:background} introduces background and related work; Section~\ref{sec:comp} discusses how TVL is currently computed, while Section~\ref{sec:case} reports the case study. Section~\ref{sec:discuss} and~\ref{sec:concl} respectively discuss design guidelines and conclusions.
Our data and code are available at https://github.com/PietroSaggese/TVL\_Study.


\section{Background and Related Work}
\label{sec:background}

\subsection{Cryptoassets: derivative and non-derivative tokens}
\label{sec:der_toks}

Cryptoassets represent and facilitate transfer of value in a Distributed Ledger Technology (DLT)~\cite{auer2024technology}. 
They can be categorized according to different factors~\cite{frowis2019detecting,di2023identification,moin2020sok}.
In our context, we distinguish between derivative and non-derivative tokens~\cite{luo2024piercing}. Non-derivative or plain tokens are those without an underlying cryptoasset.
They include native tokens such as Bitcoin, governance tokens, and non-crypto-backed (NCB) stablecoins, i.e., cryptoassets like USDC whose value is pegged to a target currency and whose reserves are not composed of cryptoassets.
Derivative tokens represent instead a receipt token that grants a claim on an underlying cryptoasset. 
They (non-exhaustively) include liquidity pool (LP) tokens~\cite{xu2023sok}, interest-bearing tokens~\cite{gudgeon2020defi,cousaert2022sok}, liquid staking tokens (LSTs)~\cite{xiong2023leverage}, and other financial products like synthetic tokens and tokenized baskets of assets. 
Derivative tokens also include crypto-backed stablecoins such as DAI.

\subsection{Total Value Locked}
\label{sec:tvl_backgr}

DeFi protocols operate on a peer-to-pool model: investors deposit their cryptoassets into accounts that pool the invested funds to offer financial services~\cite{xu2023sok,xu2022short}. 
Total Value Locked is the primary metric for assessing 
the performance of DeFi protocols and the broader DeFi ecosystem, and it is computed as the aggregate value of cryptoassets deposited in the smart contracts that constitute one protocol.

More formally, let $\mathcal{P} = \{p_1,\dots,p_n\}$  be the set of all DeFi \textbf{projects} or \textbf{protocols} and $\mathcal{A} = \{a_1,\dots,a_k\}$ be the set of all \textbf{cryptoassets} deployed on a DLT (we remove time subscripts for ease of notation). 
For one project $p$, $\mathcal{C}^{p} = \{c^p_1,\dots,c^p_m\}$ is the set of project-specific contracts. 
The association of a contract to a project must be \textit{mutually exclusive}, i.e. $\mathcal{C}^{p} \cap \hspace{0.2em} \mathcal{C}^{p'} = \emptyset$.
For each contract $c \in \mathcal{C}^p$, a vector $\mathbf{q}$ of length $k$ indicates the amount of tokens locked into the contract, and the price of each token $a \in \mathcal{A}$ is $\pi_{a}$, denominated in a common currency. We then construct the matrices

\vspace{-0.2cm}

\[
\begin{array}{ccc}
	Q^p = 
	\begin{bmatrix}
		q^p_{c_1,a_1}  & \cdots & q^p_{c_1,a_k} \\
		q_{c_2,a_1}  & \cdots & q^p_{c_2,a_k} \\
		\vdots  & \ddots & \vdots \\
		q_{c^p_m,a_1}  & \cdots & q_{c^p_m,a_k} \\
	\end{bmatrix}, \hspace{1cm}  &
	\Pi = \begin{bmatrix} \pi_{1} \\ \pi_{2} \\ \vdots \\ \pi_{k} \end{bmatrix}, \hspace{0.3cm} 
	V^p = \begin{bmatrix} v^p_1 \\ v^p_2 \\ \vdots \\ v^p_m \end{bmatrix} 
\end{array}
\]

Where $V^p = Q^p \cdot \Pi$. Then $TVL^p$, denoting the value locked in a protocol, and $TVL$, denoting the value deposited in a DLT, are respectively computed as:

\vspace{-0.2cm}

\[
TVL^p = \sum\limits_{i=1}^m v^p_i, 
\qquad
TVL = \sum\limits_{p=1}^n TVL^p
\]

\subsection{TVL aggregators}
\label{sec:aggregators}

Several aggregators publishing DeFi TVL estimations exist, e.g. DappRadar~\cite{dappradar_methods}, Stelareum~\cite{stelareum}, DeFi Pulse~\cite{defipulse_methods}, Coingecko~\cite{coingecko_methods}, and DeFiLlama~\cite{defillama}. 
The estimations they report ultimately rely on on-chain token volumes and price information, but the methods used to compute these figures and the completeness of the related documentation vary greatly. 

DeFiLlama is one of the most comprehensive aggregators in terms of TVL information disclosure~\cite{luo2024piercing}. 
Its core infrastructure, the DefiLlama-Adapters GitHub repository~\cite{defillama_adapters}, enables the community to integrate new protocols through plugins, supposedly developed by the protocols' own maintainers, and is public --- so anyone can observe how TVL is computed. 
However, since the plugins are contributed directly by the community, transparency and reliability concerns may arise.
The DeFiLlama maintainers discourage the use of data not directly sourced from cryptocurrency nodes (with the exception of exchange rates from Coingecko)~\cite{defillama_noapis}, but the framework does not impose strict limitations on off-chain data sources. 
Even when utilizing on-chain data only, the plugin creators can use self-defined, not documented on-chain functions to compute TVL, and the projects do not publish explicitly the contracts and tokens utilized to compute TVL. This information is implicit in the plugin code and no further documentation is provided by protocols. 
Moreover, the computing code may depend on specific implementations and therefore vary across time.  
There is also no description informing whether plugin contributors are actual protocol maintainers or other users.

Other aggregators provide more limited documentation, and in some cases, we could not find their TVL-computing code. Some aggregators require DeFi projects to submit a request to be included in TVL calculations, publishing only the final TVL values~\cite{stelareum,defipulse_methods}.
These differences can introduce self-selection bias, i.e. the reported TVL may differ consistently across platforms if protocols share information selectively.

Finally, some platforms publish information related to their asset selection criteria, describing what tokens are included in TVL computations~\cite{defipulse_methods,defillama}.
To date, this process is not standardized.
Notably, also DeFiLlama's framework allows to select or exclude, e.g., governance tokens, borrowed tokens, and others~\cite{coindesk_toggles}; its TVL estimations at the end of 2024 for Ethereum range from 80 to 190 bln\$.

\subsection{Related work}
\label{sec:litrev}

Surprisingly, despite the limitations discussed above and the central role TVL plays in DeFi, little research has systematically examined how TVL is computed.
This is especially important considering that several studies base their analyses on TVL, e.g., to investigate DeFi growth rates~\cite{stepanova2021review} or its relation to ETH returns~\cite{shilov4432586return}, used it as a variable in econometric~\cite{maouchi2022understanding,zhou2022sok,csoiman2022return} and machine learning models~\cite{fan2022towards}, or examined it in the context of relevant DeFi indicators~\cite{metelski2022decentralized,katona2021decentralized}.

One major limitation of TVL identified by the academic community is the double counting problem~\cite{chiu2023understanding,saengchote2021defi}.
A first solution to this issue was proposed by Luo et al.~\cite{luo2024piercing}, who focused on asset selection criteria and devised a novel metric to address double counting. \textit{Total Value Redeemable} (TVR) is defined as the value that can be ultimately redeemed from a DeFi ecosystem, i.e., the sum of non-derivative tokens deposited in DeFi protocols.

Instead, we are not aware of any prior work assessing comprehensively the reproducibility and verifiability of TVL estimates. 
Our study fills this gap and complements existing literature~\cite{luo2024piercing} by investigating empirically how TVL is effectively computed, what methodologies are used, and the extent to which TVL is reproducible using available on-chain data.


\section{TVL Computation in Practice}
\label{sec:comp}

\begin{figure}[htbp]
	\centering
	\includegraphics[width=\columnwidth]{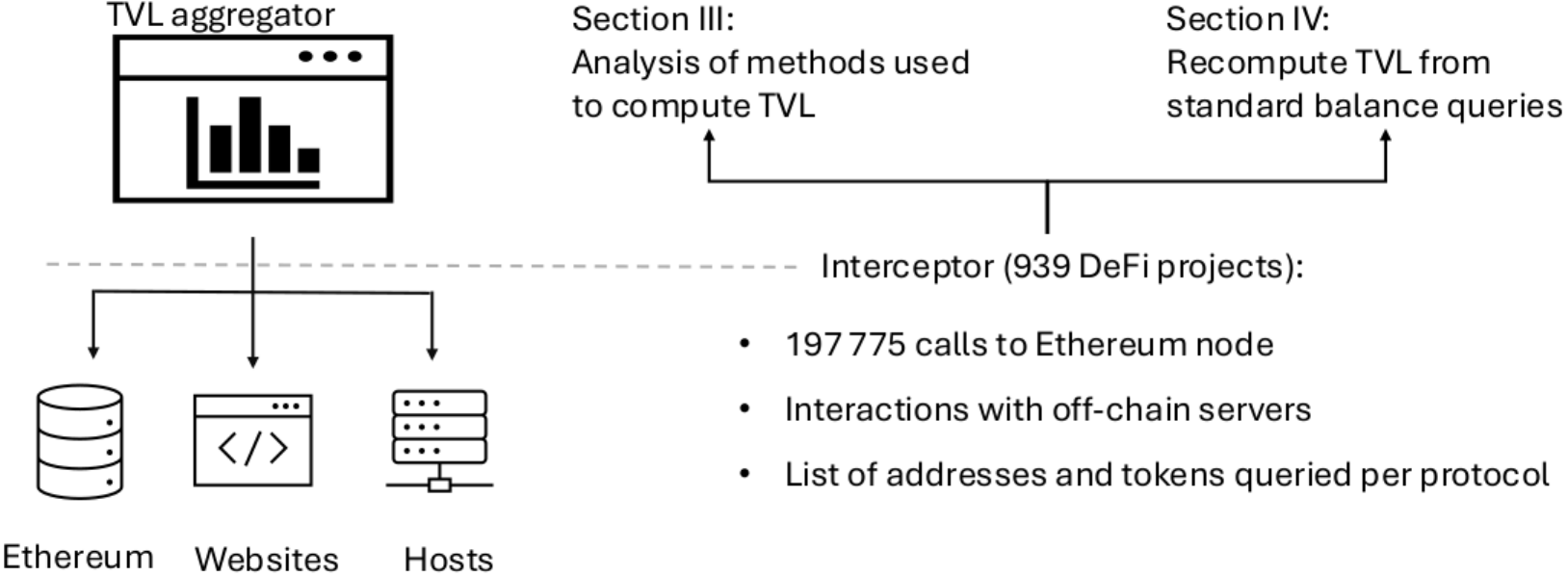}
	\caption{\textbf{Analysis pipeline.} For each protocol, we execute the plugin code provided by the TVL aggregator and record all interactions with both on-chain and off-chain sources.}
	\label{fig:pipeline1}
\end{figure}

In this section, we examine how TVL is computed in practice, what data sources are used and what potential challenges to reproducibility and standardization arise from those design decisions.
We focus on the projects listed on DeFiLlama for the reasons discussed in Subsection~\ref{sec:aggregators} (in principle, the analysis could be conducted on other aggregators publishing their TVL-computing code).
We restricted the blockchain data collection to Ethereum as it plays a major role in DeFi with 66\% of total TVL according to DeFiLlama. We expect that our findings can be generalized to most alternative chains that are EVM compatible and mirror Ethereum’s ecosystem.
The pipeline of the analysis is shown in Figure~\ref{fig:pipeline1}.

\subsection{Measurement method and summary statistics}
\label{sec:interc}

To conduct our analyses, we developed a data extraction pipeline that proxies and systematically records all interactions between the DefiLlama tooling and its runtime environment during TVL computation. 
This encompasses all calls directed towards the Ethereum node software, as well as interactions with external hosts and web services.
Our instrumentation records on-chain interactions for arbitrary combinations of Ethereum block height, commit, and chains 
(more details on this procedure are reported in Appendix~\ref{sec:implementation}).

Following the recording process, we filter and interpret the collected on-chain data to identify the functionalities used and accounts accessed per protocol. 
To interpret the data, we map the signatures of the called functions either directly when stored or  using the Ethereum Signature Database 4byte~\cite{4byte}.

Our main dataset consists of the interactions between the DefiLlama repository and its environment, collected on January 4 2024 (commit \seqsplit{6764756f9270ab6a3047c06c13c0b1b2d32a3247}). It includes $\NProjsUseEth/$ projects deployed on Ethereum out of $\NProjs/$ total projects. It comprises their interactions with external servers and the blockchain infrastructure, the latter resulting in \num{197775} proxied calls.
To validate consistency, we repeated the collection on four other dates (04 Jan 2023, 04 Apr 2023, 04 Jul 2023, 04 Oct 2023).

\begin{table}[t]
	\centering
	{\scriptsize
\begin{tabular}{l@{\hskip 4pt}r@{\hskip 8pt}r@{\hskip 10pt}l}
\toprule
Method & Count & \makecell[c]{ Num. of \\protocols } & Returns the \dots \\
\midrule
\addlinespace[3pt]
balanceOf & \num{102655} & \num{641} & \makecell[l]{\dots account balance of another account \\ (owner address)~\cite{ERC20EIP}.} \\
\specialrule{0.001pt}{1pt}{3pt}
eth\_getBalance & \num{1696} & \num{170} & \makecell[l]{\dots ETH balance of the account of \\ a  given address~\cite{eth_getbalance}.} \\
\specialrule{0.001pt}{1pt}{3pt}
totalSupply & \num{2577} & \num{154} & \makecell[l]{\dots total token supply~\cite{ERC20EIP}.} \\
\specialrule{0.001pt}{1pt}{3pt}
getReserves & \num{3016} & \num{121} & \makecell[l]{\dots reserves of token0 and token1 used to \\ price trades and distribute liquidity~\cite{UniswapDocs}.} \\
\specialrule{0.001pt}{1pt}{3pt}
token0 & \num{1040} & \num{94} & \makecell[l]{\dots address of the first pair token
~\cite{UniswapDocs}.} \\
\specialrule{0.001pt}{1pt}{3pt}
token1 & \num{1040} & \num{94} & \makecell[l]{\dots address of the second pair token
~\cite{UniswapDocs}.} \\
\specialrule{0.001pt}{1pt}{3pt}
symbol & \num{1835} & \num{64} & \makecell[l]{\dots symbol of the token.~\cite{ERC20EIP}.} \\
\specialrule{0.001pt}{1pt}{3pt}
allPairsLength & \num{46} & \num{45} & \makecell[l]{\dots total number of pairs created \\ through the factory~\cite{UniswapFactoryDocs}.} \\
\specialrule{0.001pt}{1pt}{3pt}
underlying & \num{1613} & \num{43} & \makecell[l]{\dots address of the underlying token~\cite{AaveV2ProviderContr}.} \\
\specialrule{0.001pt}{1pt}{3pt}
totalAssets & \num{236} & \num{26} & \makecell[l]{\dots total quantity of all assets under control \\ of a Vault
~\cite{VyperContr}.} \\
\bottomrule
\end{tabular}
}
	\caption{\textbf{Most frequently queried functions.} The \textit{balanceOf} method is the most commonly used in TVL computations, along with \textit{eth\_getBalance}. Other functions, such as \textit{totalAssets}, offer alternative approaches to retrieving balances.}
	\label{tab:calls}
\end{table}

Table~\ref{tab:calls} reports the most relevant functions called on-chain, ranked by number of protocols using them and complemented by a short description of their use. 
As expected, the most frequently called methods are balance queries for ERC-20 compatible tokens (\textit{balanceOf}) and Ether (\textit{eth\_getBalance}), but alternative functions exist (e.g., \textit{totalAssets}). 
\textit{BalanceOf} is primarily queried on wETH, stablecoins, governance tokens, and staked tokens (see Appendix~\ref{sec:implementation} for further details).
Other methods are used to query supplementary information, e.g., the token address or its symbol (\textit{token0}, \textit{token1}, \textit{symbol}), are used to price trades and distribute liquidity (\textit{getReserves}), or retrieve token information (\textit{underlying, totalSupply}).

\subsection{Reproducibility and reliance on off-chain data}
\label{sec:reproduc}

Relying on public and transparent on-chain information entails a number of advantages compared to off-chain data sources. The latter is easier to tamper with, creates more dependencies not in control of the user, and is less transparent, since one has limited visibility into web services' internal operations.
We thus run each protocol's computation code and investigate whether the infrastructure relies solely on on-chain data or also exploits external hosts. We also document whether the computation executes correctly or if errors occur.

\begin{figure}[t]
	\centering
	\hspace*{-0.3cm}
	\begin{tikzpicture}[scale=0.83]
		
		\draw[thick] (-4.5,-2.8) rectangle (6.5,2); 
		
		\begin{scope}[rotate=70]
			\draw[thick] (0,0) circle (1.75cm);
			\node at (0,3) {On-chain calls};
			\node at (-0.4,2.8) {produced};
			
			\draw[thick] (-1,-1.5) circle (1.1cm);
			\node at (-0.4,-4.2) {External hosts used};

			\draw[thick] (0.5,-1.7) circle (1.1cm);
			\node at (1.8,-4) {Errors are raised}; 
		\end{scope}
		
		\node at (-0.2,0.2) {729};
		\node at (1.3,0.2) {27};
		\node at (2.3,0) {62};
		\node at (1.25,-0.7) {1};
		\node at (1.8,-1) {0};
		\node at (1.2,-2) {26};
		\node at (0.6,-1.2) {72};
		\node at (-3.4,-2.5) {Others = 22};
		\node at (5.4,1.7) {\textbf{Total} = 939};
	\end{tikzpicture}
	\caption{\textbf{Projects using off-chain in addition to on-chain sources.} Projects are categorized based on their reliance on external hosts and whether any errors occurred during data collection. While for $\sioo/$ projects, on-chain calls were executed without errors and without the use of external sources, $99$ projects depend on external servers.}
	\label{fig:proj_space}
\end{figure}
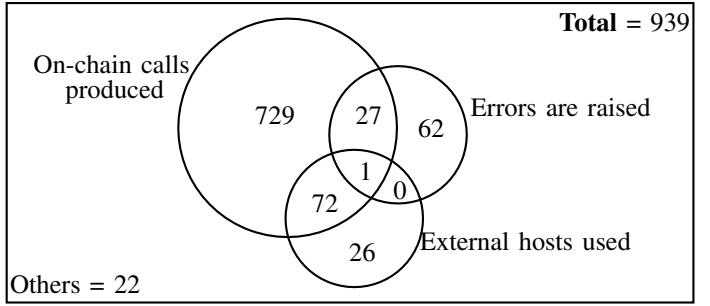

Figure~\ref{fig:proj_space} represents graphically the space of $\NProjsUseEth/$ DeFiLlama projects analyzed. 
For the largest group ($N =  \sioo/$), the procedure is executed without errors and no external hosts are used. 
While 73 projects utilized both external hosts and on-chain data sources, for  $\sooi/$ protocols we observe interactions with external hosts but no direct on-chain data could be retrieved; since no errors occurred in their collection procedure, we infer that they solely rely on external services. 
The TVL of the remaining protocols is either computed despite errors being produced during computations ($N = 28$) or not computed at all ($N = 62$). Finally, 22 projects did not produce any interaction and were discarded.

Among the 64 external servers identified, some pose a smaller threat to verifiability, e.g., TheGraph, an indexing protocol for accessing blockchain data;  others appear to be associated with third parties or specific protocols.
The errors are mostly related to technical failures such as the block height provided not being accepted, missing fields or parameters, and asynchronous operations that were not completed successfully. Further details are reported in Appendix~\ref{sec:app_repr}. 

Arguably, TVL is computable with on-chain data alone, but in practice we find several instances where other data sources are used. Despite the clarity of the DeFiLlama maintainers' objectives, certain protocols ($10.5\%$, including Uniswap V1 and V2) rely partially or entirely on data extracted from external servers for computations, partially compromising verifiability.

\subsection{Heterogeneity of on-chain interactions}
\label{sec:results_2}

Even when utilizing on-chain data only, challenges to explainability may arise. Projects use a wide set of functions alternative to standard \textit{eth\_getBalance} and \textit{balanceOf} queries to acquire on-chain balance data. These could hinder interpretation and introduce sources of tampering or inaccuracies if their logic is unclear. 
We thus focus on the \num{197775} calls directed to an Ethereum node and analyze to what extent the methods used to compute TVL are heterogeneous and the computation approach standardized across protocols.

First, to study if certain functions play an outsized role or if anomalous patterns emerge, we fit to a power-law distribution the frequency of occurrence both of the functions called and of the token calls in \textit{balanceOf} functions.
Analyzing the frequency of occurrence of the functions called, we observe that \textit{balanceOf} emerges as an extreme value with respect to the power-law fit, further highlighting its central position in TVL measurement. In general, the results of the fit are consistent with the behavior of a heavy-tailed distribution; the estimated $\alpha$ range from $1.63$ to $1.85$, coherently with findings of previous network studies on Ethereum blockchain data~\cite{kitzler2023disentangling,lee2020measurements}.

Next, we devise an approach to identify the functions alternative to \textit{balanceOf} and \textit{eth\_getBalance} (used by $78.6\%$ of protocols) whose name and description indicate that they are likely alternative functions self-defined by projects to compute TVL. We conduct a keyword-based search through regular expressions on the function signature names. We find $68$ alternative functions, used by $14.2\%$ of protocols, that plausibly contribute to TVL computation. Out of these, four functions, called $94$ times only, have name \textit{balanceOf}, but different signatures than the standard ERC-20.
The remaining protocols either rely on external hosts or exploit functions ($N = 88$) that are not clearly matched to a balance-querying functionality after a manual check. 
While standard functions are simpler to interpret, the alternative functions are specific to a few or just one project, and their logic is hard to interpret without investigating the code in depth. One illustrative example is represented by Lido, whose TVL is computed through one single function (\textit{getTotalPooledEther}) that likely returns as output the total protocol TVL.

In summary, while most protocols ($78.6\%$) use standard token balance queries, a subset of alternative functions ($N = 68$) is employed; however, this makes it harder to understand how TVL is computed for protocols utilizing such functions.
Further details on the network analysis and on the detection of the alternative functions are reported in Appendix~\ref{sec:app_het}.

\subsection{Changes in TVL computation methods}
\label{sec:commits_comp}

Another thread to verifiability and consistency are changes to the code itself. 
TVL computation relies on self-reports from DeFi protocols maintainers, and evidence of manipulations to purposedly inflate TVL exists~\cite{nelson:2022a,nuzzi:2021a}.
More generally, computations may depend on specific implementations and therefore vary across time. 
As our pipeline enables the data collection for various commits, we repeat the data gathering at four alternative dates (4th of January, April, July, and October 2023) and analyze changes in the Ethereum queries per protocol. 
We focus on \textit{balanceOf} for the interpretability of the associated calls and measure to what extent projects called different addresses and tokens over time.

\begin{figure}[t]
	\centering
	\includegraphics[width=0.75\linewidth]{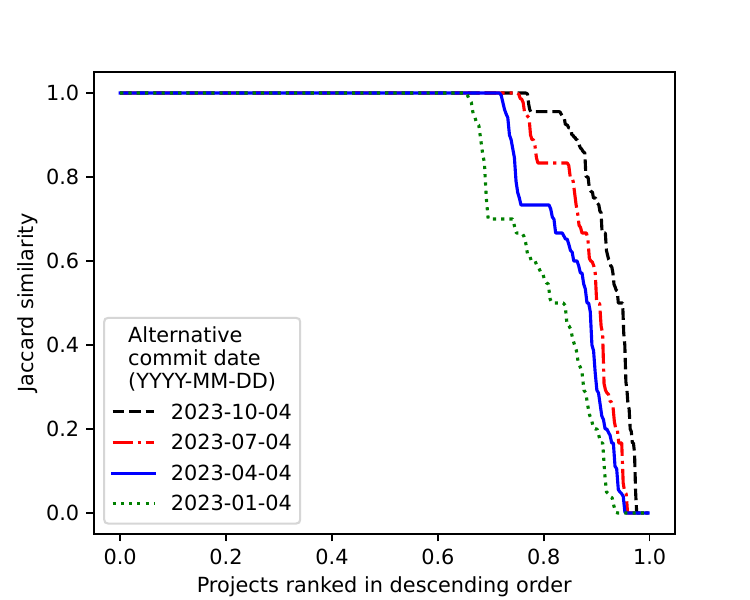}
	\caption{\textbf{Pairwise Jaccard similarity, computed for each protocol on the set of \textit{balanceOf} calls queried in the main and four older commits}.  Each line reports the values relative to one specific commit. Projects are ranked in descending order based on their similarity score (ranging from 0 to 1). A value of 1 indicates no changes among the compared sets. } 
	\label{fig:comp_main}
\end{figure}

To do so, for each protocol we compare the set of \textit{balanceOf} calls executed in the main commit (Jan 4 2024) to the set of \textit{balanceOf} calls executed in each older commit analyzed\footnote{We only compare protocols that existed at both commit dates and exclude those that used hosts or raised errors during the data collection. We note that we do not investigate commits executed earlier than 2023 because the number of projects that can be compared decreases with time (from nearly 500 between Oct 2023 and Jan 2024 to less than 300 between Jan 2023 and Jan 2024) and because our pipeline is optimized to capture data on recent implementations of the repository.}. We quantify the differences between sets as the pairwise Jaccard similarity; values range from 0 to 1, and a value of 1 indicates no changes among the compared sets. 
Figure~\ref{fig:comp_main} shows the results: each line reports the values relative to one specific commit. Projects are ranked in descending order, based on their Jaccard similarity score. 
We observe that in each older commit the set of calls does not change for most projects. The average Jaccard index ranges from $0.93\%$ (Oct 2023, black dashed line) to $0.89\%$, $0.86\%$ and $0.81\%$, respectively in Jul, Apr, Jan 2023. The Jaccard similarity decreases as expected with time, but moderately. 
Protocols typically do not modify the TVL computation code significantly across time. 
This finding has a two-fold interpretation: on the one hand, computation is relatively stable over time and thus less challenges arise from this perspective; on the other hand, it is possible that protocols are not systematically updating their plugins despite smart contract changes being implemented.

Similarly, we measure how the use of functions alternative to \textit{balanceOf} varies across commits. 
We find that the ratio between the number of alternative function queries over that of \textit{balanceOf} and \textit{eth\_getBalance} calls is higher in older commits ($28.2\%$ in Jan 2023, $22.7\%$ in Apr 2023, $19.9\%$ in Jul 2023, and $12.2\%$ in Oct 2023, against $8.9\%$ in Jan 2024). We interpret this as a sign that heterogeneity in computation methods has reduced over time. 
Appendix~\ref{sec:app_changes} reports additional data, such as the evolution in time of the number of standard and non-standard calls, the full identifiers of older commits, and additional results, including the Jaccard similarity scores on all protocols and in absolute numbers.

\subsection{Equivalent balance queries linked to multiple protocols}
\label{sec:dc}

Double counting represents a major issue in the computation of financial metrics like TVL. It occurs when the value of an asset is counted more than once, leading to an inflated representation of total assets within the system.
Previous research~\cite{luo2024piercing} introduced the Total Value Redeemable as an alternative to TVL and partly addressed the problem by counting the value of non-derivative tokens only.
From a technical perspective, double counting can also occur if a smart contract included in the TVL calculation is not exclusively associated with one project, causing the corresponding TVL to be counted multiple times.

We thus search for balance queries that are not exclusively linked to a single protocol, thus potentially leading to such double counting. Notably, we identify $230$ \textit{balanceOf} and $10$ \textit{eth\_getBalance} calls that appear to be associated with different protocols on the same input smart contract and for the same token address.
A possible explanation for this finding is that these contracts are managing interactions across protocols\footnote{While we assume that balances queried during computation directly contribute to the projects TVL, we acknowledge the possibility that they don't.}. We measure and discuss the magnitude of this phenomenon in economic terms in the next section. 
Appendix~\ref{sec:app_dc} reports a full list of the addresses and tokens involved.


\section{Case study: TVL Reconstruction From On-chain Data and vTVL}
\label{sec:case}

In principle, TVL can be calculated entirely on public, immutable blockchain information: this may provide a solution to some  of the challenges emerged in TVL computation affecting its reproducibility and verifiability.
Building on this, we conduct a case study to assess to what extent the TVL of individual projects can be recomputed and verified solely relying on on-chain data and standard balance queries. 
We call the resulting metric the verifiable Total Value Locked (vTVL) of a DeFi project. 
This serves as a starting point for discussing a set of recommendations and standardization attempts to improve TVL computation.

\subsection{TVL reconstruction approach}
\label{sec:res_tvl}

Having access to the functionalities and accounts queried on-chain through the DeFiLlama infrastructure, we can acquire a set of addresses per protocol containing deposited assets and provide further insights on the TVL associated with each protocol, solely relying on blockchain data (assuming that addresses called during computation contribute to the project TVL).
To have a homogeneous and standardized representation, we focus on \textit{eth\_getBalance} and \textit{balanceOf} queries. We obtain a list of addresses contributing towards the locked value for each project, along with a compilation of tokens ($N = \num{12246}$) in which all projects hold value.
For each protocol, we extract cryptoassets quantities and prices on-chain. We query the state of protocol-specific addresses for their associated tokens and extract historical monthly balance information from Jan $1^{st}$ 2021 to Feb  $1^{st}$ 2024. We price tokens by extracting exchange rate information from Uniswap V2 DEX liquidity pools~\cite{heimbach2021behavior}; in total, using this approach, we could price $942$ tokens.

Having granular information on tokens deposited into contracts, we can investigate TVL composition and increase control over selected assets.  
We categorize tokens following the distinction into non-derivative and derivative tokens~\cite{luo2024piercing} and distinguish seven categories: Ether and its wrapped token wETH, wrapped BTC (wBTC), non-crypto-backed stablecoins, crypto-backed stablecoins, governance tokens, derivative tokens, and others.
We note that we report separately the balance queries that, as discussed in Subsection~\ref{sec:dc}, are non-exclusively associated with one protocol and therefore potentially contribute to double counting, as we could not disentangle which project they should be associated with.

We thus recompute the value held by each protocol and call these estimates the verifiable Total Value Locked (vTVL).
For comparison, we also download historical TVL values\footnote{The API data are reported by DeFiLlama distinguished by chain and type. We include all values associated with the Ethereum blockchain (columns `Ethereum', `Ethereum-borrowed', `Ethereum-pool2', `Ethereum-staking', `Ethereum-vesting').} published per protocol from the DeFiLlama API service~\cite{defillama_apis}. 
We compute for each project the \textit{Discrepancy Ratio}, defined as the average ratio of the vTVL estimations over the data posted through the DeFiLlama APIs, normalized by subtracting one. A value of zero indicates perfect correspondence, while $-1$ indicates that vTVL equals zero.
To interpret this metric, we recall that our estimations are an underestimation rather than an overestimation of the reported figures. 
We stress that the discrepancies should not be interpreted as a measure that protocols are inflating values; rather, they indicate to what extent we are able to independently verify the reported figures.
Further details on this are given in Appendix~\ref{sec:app_reconstr}.

\begin{figure*}[h!]
	\centering
	\begin{subfigure}[b]{0.4\textwidth}
		\centering
		\includegraphics[width=\textwidth]{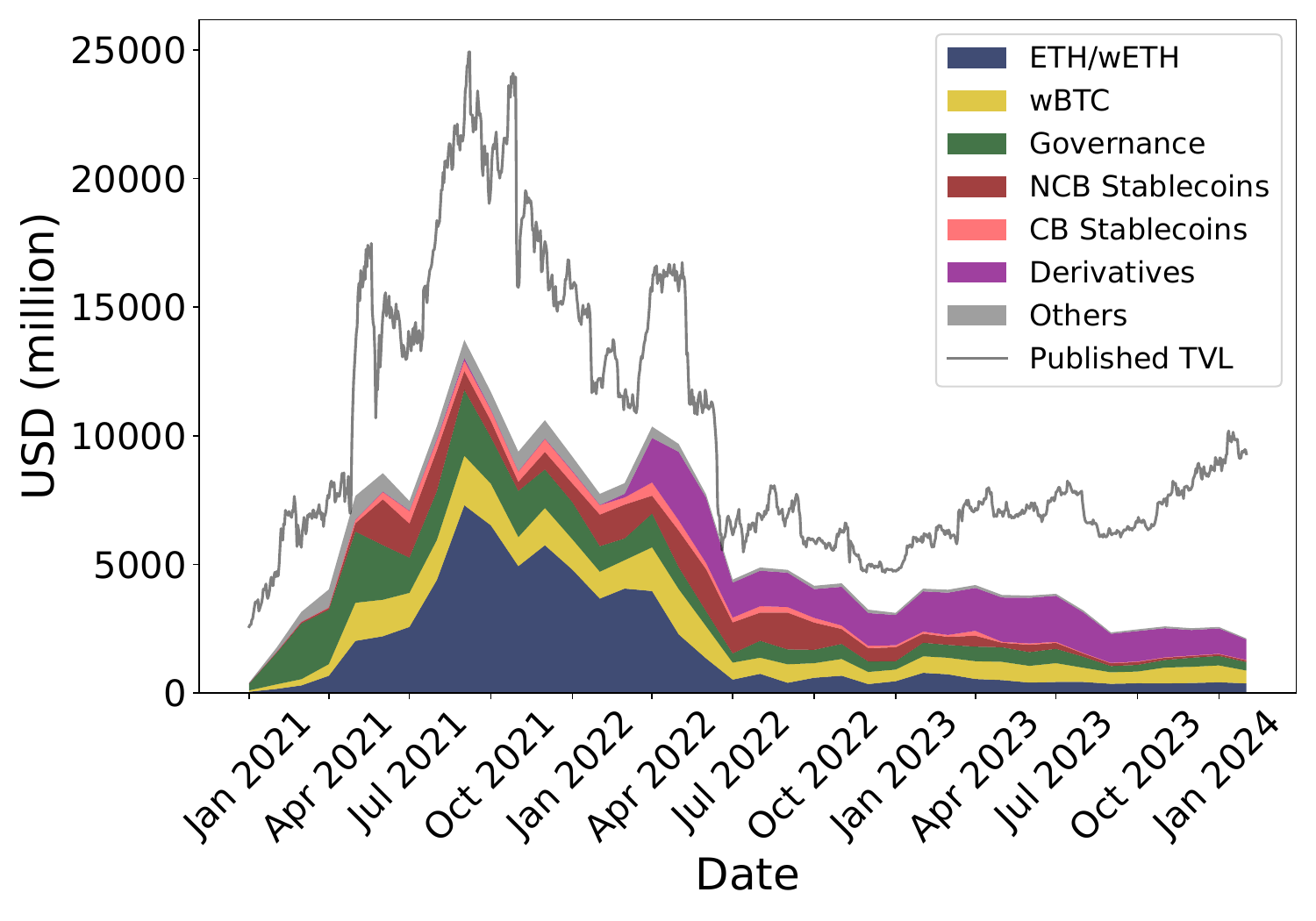}
		\vspace{-0.7cm}
		\caption{Aave (v2 and v3)}
		\label{fig:aave2_stack}
	\end{subfigure}
	\hspace{0.7cm}
	\begin{subfigure}[b]{0.4\textwidth}
		\centering
		\includegraphics[width=\textwidth]{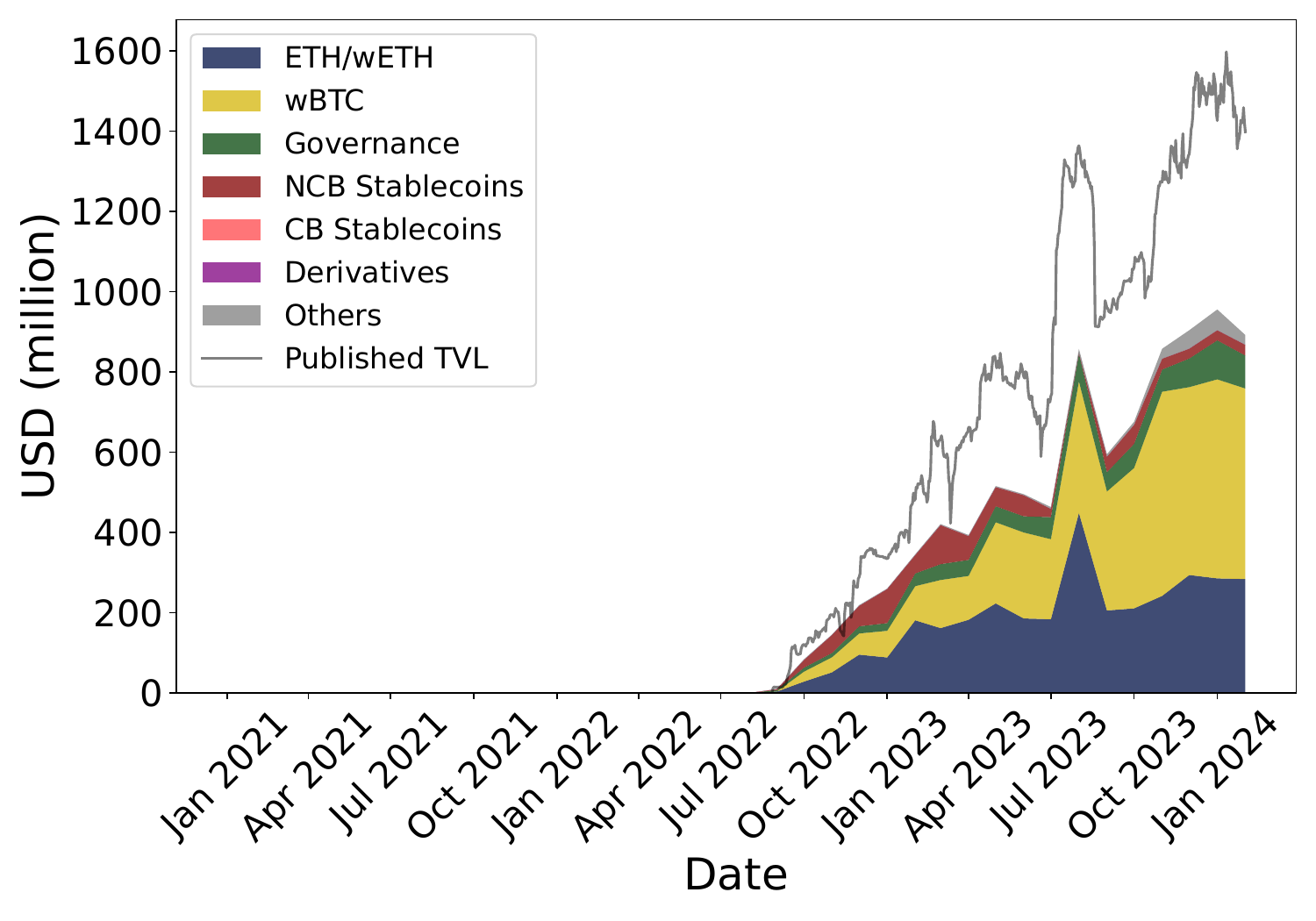}
		\vspace{-0.7cm}
		\caption{Compound-v3}
		\label{fig:compound_stack}
	\end{subfigure}
	\begin{subfigure}[b]{0.4\textwidth}
		\centering
		\includegraphics[width=\textwidth]{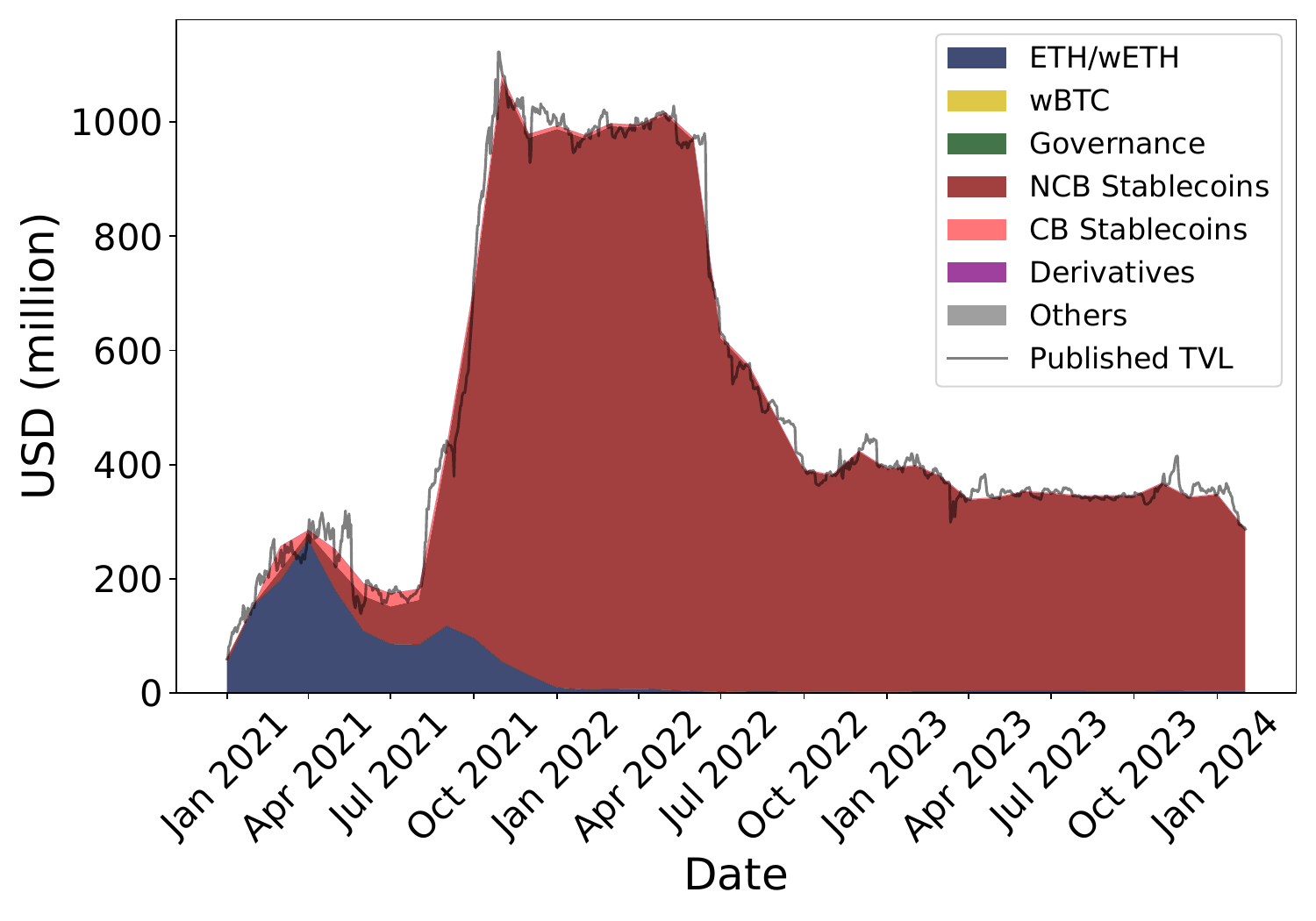}
		\vspace{-0.7cm}
		\caption{dYdX}
		\label{fig:dydx_stack}
	\end{subfigure}
	\hspace{0.7cm}
	\begin{subfigure}[b]{0.4\textwidth}
		\centering
		\includegraphics[width=\textwidth]{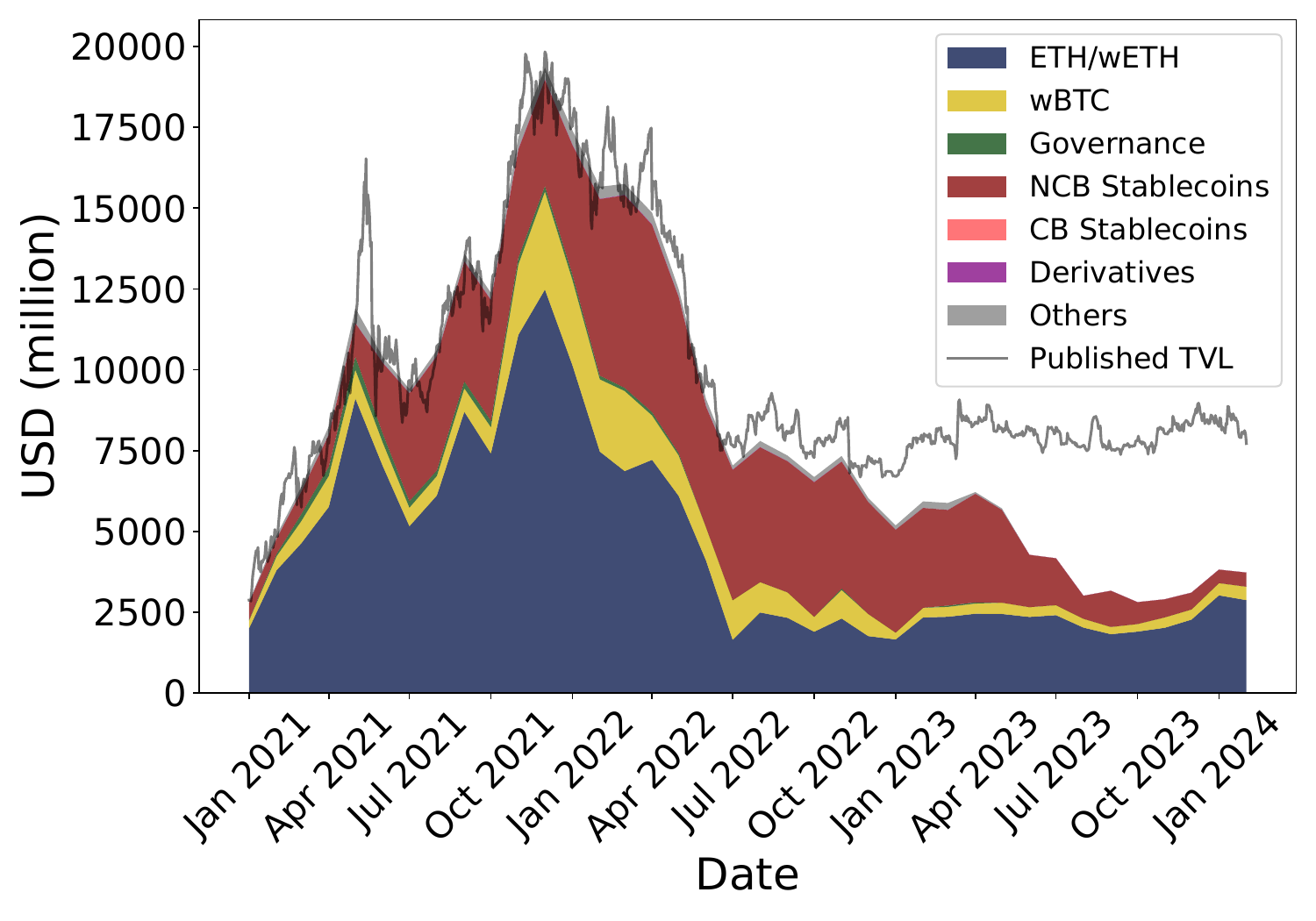}
		\vspace{-0.7cm}
		\caption{Maker}
		\label{fig:maker_stack}
	\end{subfigure}
	\begin{subfigure}[b]{0.4\textwidth}
		\centering
		\includegraphics[width=\textwidth]{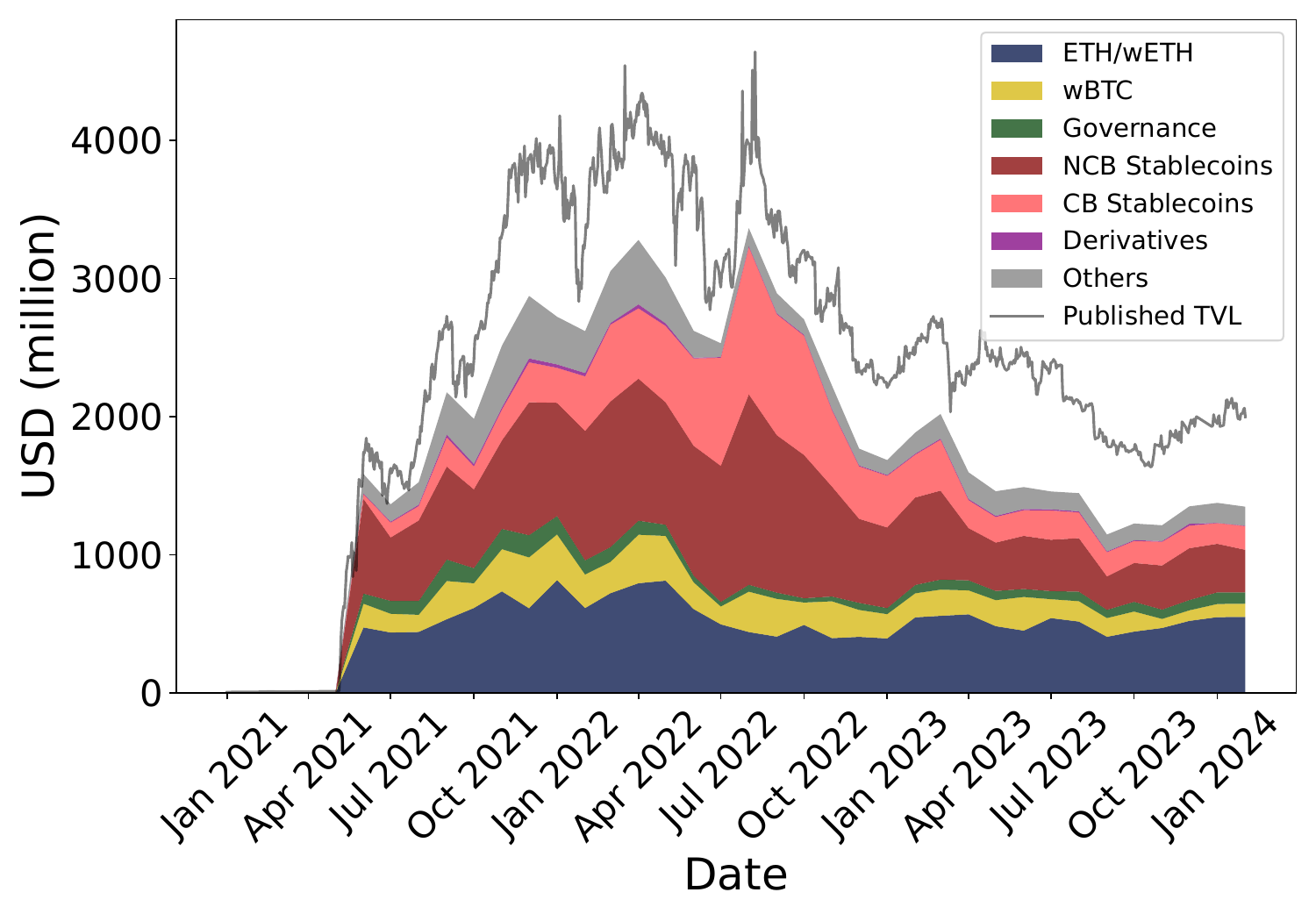}
		\vspace{-0.7cm}
		\caption{Uniswap-v3}
		\label{fig:uniswap_stack}
	\end{subfigure}
	\hspace{0.7cm}
	\begin{subfigure}[b]{0.4\textwidth}
		\centering
		\includegraphics[width=\textwidth]{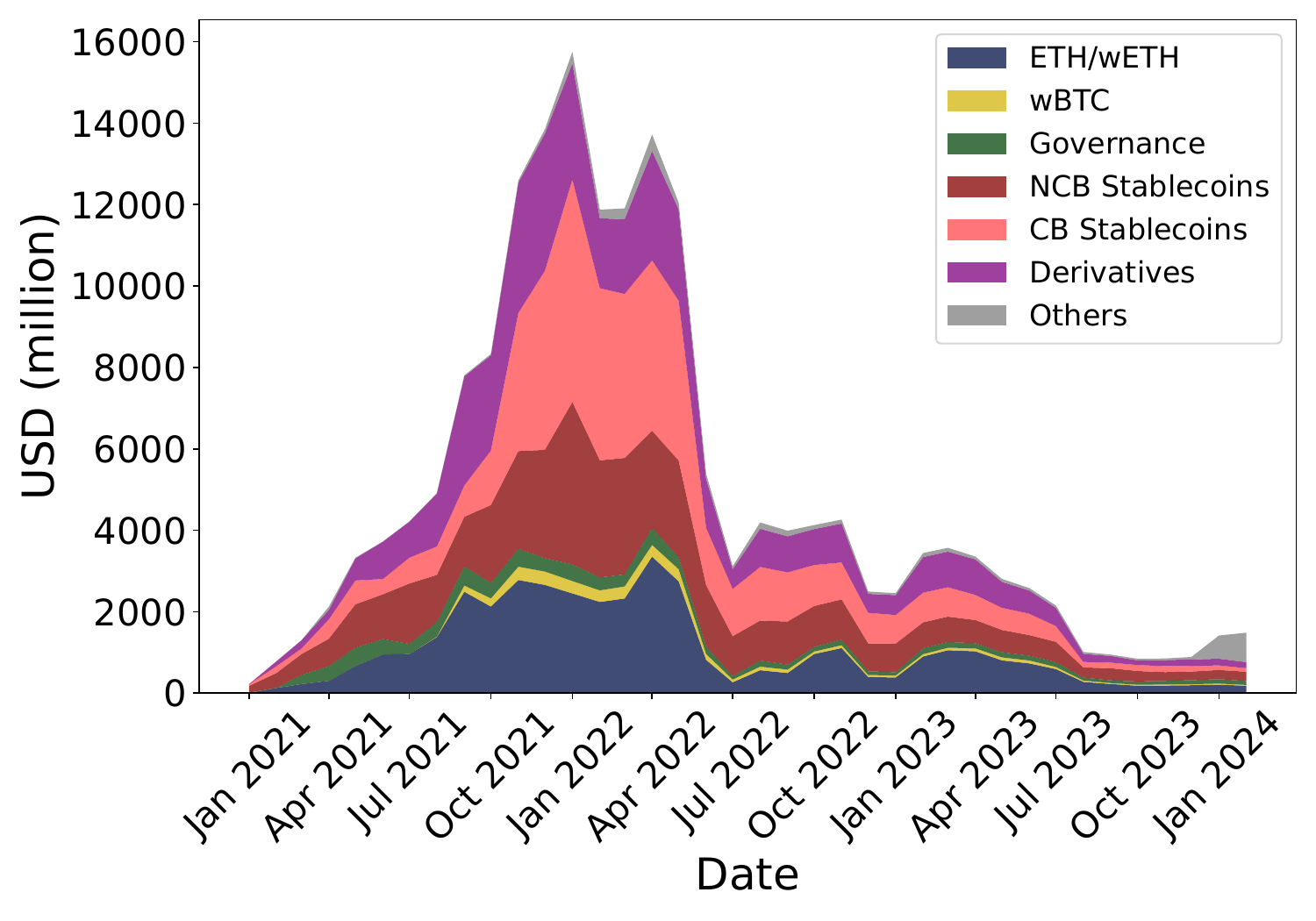}
		\vspace{-0.7cm}
		\caption{}
		\label{fig:dc_stack}
	\end{subfigure}
	\caption{\textbf{Verifiable Total Value Locked (vTVL).} We query on-chain information for relevant DeFi protocols --- Aave v2 and v3~(\subref{fig:aave2_stack}), Compound v3~(\subref{fig:compound_stack}), dYdX~(\subref{fig:dydx_stack}), Maker~(\subref{fig:maker_stack}), Uniswap v3~(\subref{fig:uniswap_stack}) --- and compare it to published off-chain TVL data. Each plot represents the evolution in time of their vTVL as a stacked plot, divided in seven categories: Ether and wETH, wBTC, governance tokens, non-crypto-backed stablecoins, crypto-backed stablecoins, and uncategorized tokens. 
    Panel~(\subref{fig:dc_stack}) shows instead the evolution in time of the value of $240$ \textit{balanceOf} and \textit{get\_ethBalance} queries called on the same contracts and tokens but associated with different protocols (see Subsection~\ref{sec:dc}), potentially contributing to double counting.}
	\label{fig:stacks}
\end{figure*}

\subsection{Case study: analysis and results} 

We analyze $400$ protocols with at least one \textit{eth\_getBalance} or \textit{balanceOf} recorded call and for which we could gather off-chain API data. 
Panels~(\subref{fig:aave2_stack}) to (\subref{fig:uniswap_stack}) of Figure~\ref{fig:stacks} show the results for five protocols selected for their relevance in the DeFi ecosystem: in order, Aave (v2 and v3), Compound-v3, dYdX, Maker, and Uniswap-v3. Each panel shows a stacked plot of the vTVL divided by token categories as discussed above. The black line represents the (total) TVL value published on DeFiLlama. 
We observe that vTVL is mostly composed of ETH/wETH, wBTC, and non-crypto-backed stablecoins. 
In most cases, the data reconstructed from on-chain and off-chain data are consistent; for some projects, the data match almost perfectly, while for others, we observe a partial discrepancy, but their overall trend is consistent.
The discrepancies can arise for varying reasons, e.g., the use of alternative methods to compute balances, the reliance on external hosts to produce values (e.g. Maker) or the presence of errors during the interception procedure (e.g. Uniswap V3), but also for the lack of price data for certain tokens. Further details are reported in Appendix~\ref{sec:app_reconstr} and limitations are discussed in Section~\ref{sec:discuss}.

Panel~(\subref{fig:dc_stack}) reports instead the evolution over time of the value held in the contract accounts identified in Subsection~\ref{sec:dc} for the balance queries that are repeated over multiple protocols, thus potentially contributing to double counting. Notably, the value reached a peak of almost 16bln\$ at the end of 2021: double counting is a threat to a correct interpretation of the TVL metric also at the infrastructure level, and its potential impact is economically relevant.

\begin{figure}[h]
\centering
\includegraphics[width=1.1\linewidth]{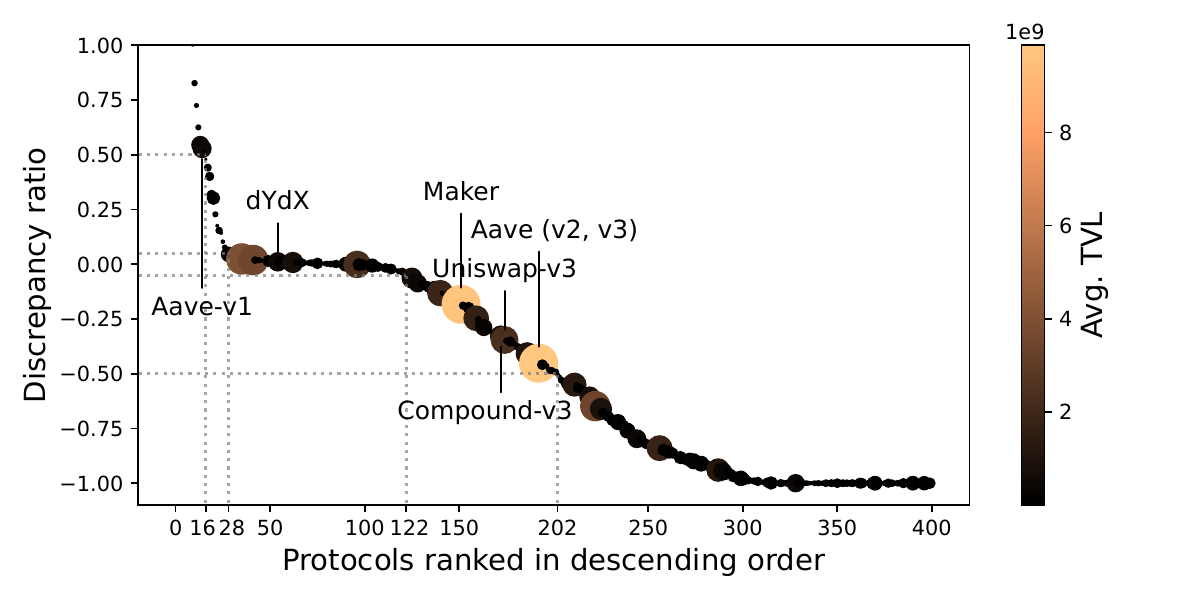}
\caption{\textbf{Differences between vTVL and  published TVL}.  Projects are dots ranked in descending order based on the \textit{Discrepancy ratio}, i.e., the ratio of on-chain estimations to off-chain data, normalized by subtracting one.}
\label{fig:pct}
\end{figure}

Finally, to obtain a broader understanding of the entire ecosystem, we compute for each project the \textit{Discrepancy Ratio}. Figure~\ref{fig:pct} shows the results. Each dot represents a project and its size is proportional to the amount of TVL they hold (according to off-chain estimations). Projects are ranked in descending order with respect to the Discrepancy ratio. The ones shown in Figure~\ref{fig:stacks} are labeled. 
For $94$ projects ($23.5\%$), the difference lies within $\pm0.05$, indicating almost perfect consistency, while for $186$ projects ($46.5\%$), the difference lies within $\pm0.5$, indicating that figures are aligned but discrepancies exist. In $98$ cases the ratio is either larger than $1$ (top left) or equal to $-1$ (bottom right), indicating large discrepancies. The latter are mostly small and less relevant protocols.

In summary, the case study shows that with vTVL, which reproduces TVL using solely on-chain data and interpretable balance queries, we can independently verify just a part of the reported TVL. Discrepancies between our estimations and published data are close to zero for about one quarter of protocols ($23.5\%$) and are aligned for about half ($46.5\%$). 
Concurrently, the study shows that it is possible to follow a more transparent approach for computing TVL. The approach we used for vTVL removes potential off-chain sources of tampering, minimizes heterogeneity in computation methods, reveals patterns that can only be investigated on-chain, like the association of one contract to multiple protocols, and enables control over the selected assets, ultimately improving reproducibility, verifiability, transparency and interpretability.

\section{Discussion: Towards TVL Standardization} 
\label{sec:discuss}

TVL today remains largely non-standardized and its computation open source to third parties. Against this backdrop, informed by the challenges identified in TVL computation and the case study conducted, we discuss a series of recommendations that may help guide the design of reproducible, verifiable, transparent and interpretable TVL estimations.

First, \textbf{TVL can and should be computed from available on-chain sources}. As discussed in Sections~\ref{sec:comp} and~\ref{sec:case}, relying on external services may hinder verifiability and reproducibility, raise concerns about transparency and create opportunities for manipulation or double counting.
Furthermore, third parties need to \textbf{know explicitly what protocol-specific smart contracts and associated tokens} are utilized to count value. This also allows to verify if the association of a contract to a project is unique or a potential source of double counting.

The use of \textbf{standard balance methods should be preferred over custom functions whenever possible}, in order to standardize computation methods and enhance interpretability of the metric. 
As discussed in Section~\ref{sec:interc}, the most common tokens are ERC-20 compliant and therefore it is straightforward to understand how their balance is quantified. 
If tokens are non-standard or rely on alternative functions, proper documentation is required to interpret results.
A promising finding in this sense is the increased predominance over time of standard functions, which revealed a trend towards homogenization.

We propose that aggregators \textbf{publish their token categorizations} and enable third parties to independently \textbf{select which assets to include or exclude in computations}, with particular attention to derivative and non-derivative tokens.
While aggregators today allow to include or exclude certain tokens from computations, such as governance tokens, borrowed tokens in lending protocols, and others~\cite{coindesk_toggles}, we could not find a detailed list of what assets are included in each category.
Furthermore, as indicated in previous research~\cite{luo2024piercing}, derivative tokens are responsible for double counting, and this aspect needs to be taken into account.
In our case study, we grouped tokens into categories and identified derivative tokens to provide deeper insights on TVL composition, allowing direct control on the assets selected for computation.

More broadly, it is also necessary to \textbf{define common standards for protocol selection criteria} and maintain a public list with included protocols. According to DeFiLlama APIs, the protocols we analyzed are grouped into 39 categories: the most common ones are DEXs, Yield, Lending, and Services (respectively $N = 78$, $68$, $55$, $34$). Notably, $N=28$ are labeled as CEXs. It is also arguable whether other categories belong to DeFi (Gaming, Oracle, Wallets) or what their purpose is (SoFi, Launchpad, Chain).
Furthermore, relying on self-reporting also implies self-selection bias to a certain extent: some protocols might not be interested in reporting data to all TVL aggregators and viceversa. 
Related to protocol selection, it is important to \textbf{complement protocol-specific information with metadata for version management}.
From the analysis in Section~\ref{sec:res_tvl}, it emerged that protocols plugins do not manage versions consistently. As protocols often deploy new versions, it is important to have a clear vision of what funds are deposited in each protocol version.

We acknowledge that future research might address some limitations of our study. 
First, our work is limited to the DeFiLlama infrastructure and to Ethereum alone, but DeFi is spread across DLTs; whilst we believe our findings can be generalized to other blockchains that replicate Ethereum, future analyses could pay specific attention to bridges and multi-chain protocols, and include other blockchains like Solana. Second, alternative computation methods to standard ones should be investigated in detail, to understand what functionalities they offer and why they are used.
Third, the token categorization needs improvement. Currently, there is no comprehensive method for identifying derivative tokens, nor a clear and shared definition. 
Similarly, the collection of token prices can be improved. We could not price all tokens in our dataset with our approach (but we did cover all the most relevant ones), and on-chain prices should be extracted from multiple DEXs and liquidity pools to account for low-liquidity scenarios. The price of certain tokens, like NFTs, might need to be complemented with exchange market price data. 
Fourth, relying only on on-chain sources is likely more costly and less efficient computationally. This approach might face resistance due to the varied nature of the platforms involved. Following the suggested approach, protocols might have to disclose more information than they do today. As this comes at a gain in terms of transparency and verifiability, it is important to further investigate how to balance these aspects.

Beyond our analysis, we note that TVR~\cite{luo2024piercing} is devised for application on an entire ecosystem and therefore helps standardize TVL at the ecosystem level.  However, it remains still unclear how to conduct asset selection at the protocol level. 
Appendix~\ref{sec:app_tvr} provides further analyses on TVL composition changes in relation to TVR and across protocol category, size, and time.
Further research should focus on this aspect.

Finally, we remark that the verifiability and transparency of TVL computations is part of a broader discussion on DeFi-related risks and issues. 
Indeed, DeFi automation and the removal of human involvement has introduced or heightened certain risks, e.g. diminishing oversight and control, or paving the way for new types of intermediaries~\cite{carter2021defi}. 
Just as TVL should be verifiable, proof-of-reserve systems are essential for stablecoins and cryptoasset trading platforms~\cite{eichengreen2023stablecoin,saggese2024assessing}.
Moreover, the need for governance makes some degree of centralization unavoidable~\cite{doerr2021defi}, and structural aspects of the system contribute to the concentration of power in the hands of few developers~\cite{fracassi2024decentralized, kitzler2024governance}.


\section{Conclusions}
\label{sec:concl}

In this work, we conduct a systematic study on 939 DeFi projects deployed in Ethereum.
We first provide a comprehensive understanding of the methodologies currently used for TVL computation; next, we examine the extent to which TVL is reproducible and verifiable using available on-chain data by introducing a new metric, the verifiable Total Value Locked (vTVL). Informed by these analyses, we propose design guidelines and possible standardization attempts in the field.

TVL is a fundamental financial metric in the DeFi ecosystem. Reaching common standards and publishing verifiable figures is critical to obtaining a clear overview of the true dimensions of the DeFi ecosystem, informing correctly users' investment decisions, and guaranteeing fair competition across protocols. 
This also supports the need for greater financial transparency and accountability to address the expectations of diverse stakeholder groups, including users, investors, and regulators.
This work provides several insights in this direction.

\section*{Acknowledgments}

The Complexity Science Hub researchers were partially funded by the Austrian security research program KIRAS of the Federal Ministry of Finance (BMF) under the project DeFiTrace (grant agreement number 905300) and the FFG BRIDGE project AMALFI (grant agreement number 898883).

\bibliographystyle{IEEEtran}
\bibliography{IEEEabrv,bibliography/references}

\begin{thebibliography}{10}
\providecommand{\url}[1]{#1}
\csname url@samestyle\endcsname
\providecommand{\newblock}{\relax}
\providecommand{\bibinfo}[2]{#2}
\providecommand{\BIBentrySTDinterwordspacing}{\spaceskip=0pt\relax}
\providecommand{\BIBentryALTinterwordstretchfactor}{4}
\providecommand{\BIBentryALTinterwordspacing}{\spaceskip=\fontdimen2\font plus
\BIBentryALTinterwordstretchfactor\fontdimen3\font minus
  \fontdimen4\font\relax}
\providecommand{\BIBforeignlanguage}[2]{{%
\expandafter\ifx\csname l@#1\endcsname\relax
\typeout{** WARNING: IEEEtran.bst: No hyphenation pattern has been}%
\typeout{** loaded for the language `#1'. Using the pattern for}%
\typeout{** the default language instead.}%
\else
\language=\csname l@#1\endcsname
\fi
#2}}
\providecommand{\BIBdecl}{\relax}
\BIBdecl

\bibitem{auer2024technology}
R.~Auer, B.~Haslhofer, S.~Kitzler, P.~Saggese, and F.~Victor, ``The technology
  of decentralized finance (defi),'' \emph{Digital Finance}, vol.~6, no.~1, pp.
  55--95, 2024.

\bibitem{defillama_adapters}
{DeFiLlama}, ``{GitHub repository - DeFiLlama-Adapters},'' available at:
  \url{https://github.com/DefiLlama/DefiLlama-Adapters}.

\bibitem{defillama_noapis}
------, ``{Why we don't accept APIs},'' available at:
  \url{https://github.com/DefiLlama/DefiLlama-Adapters/discussions/432}.

\bibitem{nelson:2022a}
D.~Nelson and T.~Wang, ``Master of anons: How a crypto developer faked a defi
  ecosystem,''
  \url{https://www.coindesk.com/layer2/2022/08/04/master-of-anons-how-a-crypto-developer-faked-a-defi-ecosystem/}.

\bibitem{chiu2023understanding}
J.~Chiu, T.~Koeppl, H.~Yu, and S.~Zhang, ``Understanding defi through the lens
  of a production-network model,'' Bank of Canada, Tech. Rep., 2023.

\bibitem{coindesk_toggles}
Coindesk, ``Data provider defillama de-emphasizes double-counted crypto
  deposits after saber revelation,'' 2022, available at:
  \url{https://www.coindesk.com/business/2022/08/05/data-provider-defillama-de-emphasizes-double-counted-crypto-deposits-after-saber-revelation/}.

\bibitem{nuzzi:2021a}
L.~Nuzzi, A.~L. Calvez, and K.~Waters, ``Understanding total value locked
  (tvl),'' 2021, available at:
  \url{https://coinmetrics.substack.com/p/coin-metrics-state-of-the-network-0c0}.

\bibitem{luo2024piercing}
Y.~Luo, Y.~Feng, J.~Xu, and P.~Tasca, ``Piercing the veil of tvl: Defi
  reappraised,'' in \emph{International Conference on Financial Cryptography
  and Data Security}.\hskip 1em plus 0.5em minus 0.4em\relax Springer, 2025.

\bibitem{frowis2019detecting}
M.~Fr{\"o}wis, A.~Fuchs, and R.~B{\"o}hme, ``Detecting token systems on
  ethereum,'' in \emph{International Conference on Financial Cryptography and
  Data Security}.\hskip 1em plus 0.5em minus 0.4em\relax Springer, 2019, pp.
  93--112.

\bibitem{di2023identification}
M.~Di~Angelo and G.~Salzer, ``Identification of token contracts on ethereum:
  standard compliance and beyond,'' \emph{International Journal of Data Science
  and Analytics}, vol.~16, no.~3, pp. 333--352, 2023.

\bibitem{moin2020sok}
A.~Moin, K.~Sekniqi, and E.~G. Sirer, ``Sok: A classification framework for
  stablecoin designs,'' in \emph{International Conference on Financial
  Cryptography and Data Security}.\hskip 1em plus 0.5em minus 0.4em\relax
  Springer, 2020, pp. 174--197.

\bibitem{xu2023sok}
J.~Xu, K.~Paruch, S.~Cousaert, and Y.~Feng, ``Sok: Decentralized exchanges
  (dex) with automated market maker (amm) protocols,'' \emph{ACM Computing
  Surveys}, vol.~55, no.~11, pp. 1--50, 2023.

\bibitem{gudgeon2020defi}
L.~Gudgeon, S.~Werner, D.~Perez, and W.~J. Knottenbelt, ``Defi protocols for
  loanable funds: Interest rates, liquidity and market efficiency,'' in
  \emph{Proceedings of the 2nd ACM Conference on Advances in Financial
  Technologies}, 2020, pp. 92--112.

\bibitem{cousaert2022sok}
S.~Cousaert, J.~Xu, and T.~Matsui, ``Sok: Yield aggregators in defi,'' in
  \emph{2022 IEEE International Conference on Blockchain and Cryptocurrency
  (ICBC)}.\hskip 1em plus 0.5em minus 0.4em\relax IEEE, 2022, pp. 1--14.

\bibitem{xiong2023leverage}
X.~Xiong, Z.~Wang, X.~Chen, W.~Knottenbelt, and M.~Huth, ``Leverage staking
  with liquid staking derivatives (lsds): Opportunities and risks,''
  \emph{arXiv preprint arXiv:2401.08610}, 2023.

\bibitem{xu2022short}
T.~A. Xu and J.~Xu, ``A short survey on business models of decentralized
  finance (defi) protocols,'' in \emph{International Conference on Financial
  Cryptography and Data Security}.\hskip 1em plus 0.5em minus 0.4em\relax
  Springer, 2022, pp. 197--206.

\bibitem{dappradar_methods}
{DappRadar}, ``{DeFi Rankings},'' available at:
  \url{https://docs.dappradar.com/rankings/defi-rankings}.

\bibitem{stelareum}
{Stelareum}, ``{Total Value Locked (TVL) in DeFi protocols},'' available at:
  \url{https://www.stelareum.io/en/defi-tvl.html}.

\bibitem{defipulse_methods}
{DeFiPulse}, ``{DeFiPulse Total Value Locked (TVL) Methodology},'' available
  at: \url{https://docs.defipulse.com/methodology/tvl}.

\bibitem{coingecko_methods}
{Coingecko}, ``{What Total Value Locked (Tvl) and Why Users Monitor This
  Metric},'' see: \url{https://www.coingecko.com/learn/total-value-locked}.

\bibitem{defillama}
{DeFiLlama}, ``{DeFiLlama website},'' available at:
  \url{https://defillama.com/}.

\bibitem{stepanova2021review}
V.~Stepanova and I.~Eri{\c{n}}{\v{s}}, ``Review of decentralized finance
  applications and their total value locked,'' \emph{TEM Journal}, vol.~10,
  no.~1, p. 327, 2021.

\bibitem{shilov4432586return}
K.~Shilov and A.~V. Zubarev, ``Return factors of ether cryptocurrency: On chain
  metrics and defi,'' \emph{Available at SSRN 4432586}.

\bibitem{maouchi2022understanding}
Y.~Maouchi, L.~Charfeddine, and G.~El~Montasser, ``Understanding digital
  bubbles amidst the covid-19 pandemic: Evidence from defi and nfts,''
  \emph{Finance Research Letters}, vol.~47, p. 102584, 2022.

\bibitem{zhou2022sok}
L.~Zhou, X.~Xiong, J.~Ernstberger, S.~Chaliasos, Z.~Wang, Y.~Wang, K.~Qin,
  R.~Wattenhofer, D.~Song, and A.~Gervais, ``Sok: Decentralized finance (defi)
  attacks,'' in \emph{2023 IEEE Symposium on Security and Privacy (SP)}.\hskip
  1em plus 0.5em minus 0.4em\relax IEEE, 2023, pp. 2444--2461.

\bibitem{csoiman2022return}
F.~{\c{S}}oiman, G.~Dumas, and S.~Jimenez-Garces, ``The return of (i) defix,''
  \emph{arXiv preprint arXiv:2204.00251}, 2022.

\bibitem{fan2022towards}
S.~Fan, T.~Min, X.~Wu, and C.~Wei, ``Towards understanding governance tokens in
  liquidity mining: a case study of decentralized exchanges,'' \emph{World Wide
  Web}, pp. 1--20, 2022.

\bibitem{metelski2022decentralized}
D.~Metelski and J.~Sobieraj, ``Decentralized finance (defi) projects: A study
  of key performance indicators in terms of defi protocols’ valuations,''
  \emph{International Journal of Financial Studies}, vol.~10, no.~4, p. 108,
  2022.

\bibitem{katona2021decentralized}
T.~Katona, ``Decentralized finance: the possibilities of a blockchain “money
  lego” system,'' \emph{Financial and Economic Review}, vol.~20, no.~1, pp.
  74--102, 2021.

\bibitem{saengchote2021defi}
K.~Saengchote \emph{et~al.}, ``Where do defi stablecoins go? a closer look at
  what defi composability really means,'' \emph{Available at SSRN 3893487},
  2021.

\bibitem{4byte}
{4byte}, ``{Ethereum Signature Database},'' see
  \url{https://www.4byte.directory}.

\bibitem{ERC20EIP}
{Ethereum.org}, ``{ERC-20: Token Standard Improvement Proposal},'' see
  \url{https://eips.ethereum.org/EIPS/eip-20}.

\bibitem{eth_getbalance}
------, ``{Ethereum JSON-RPC API},'' see
  \url{https://ethereum.org/en/developers/docs/apis/json-rpc/#eth_getbalance}.

\bibitem{UniswapDocs}
{Uniswap}, ``{Functionalities documentation},'' see
  \url{https://docs.uniswap.org/contracts/v2/reference/smart-contracts/pair}.

\bibitem{UniswapFactoryDocs}
------, ``{Factory contract documentation},'' see
  \url{https://docs.uniswap.org/contracts/v2/reference/smart-contracts/factory}.

\bibitem{AaveV2ProviderContr}
{Etherscan}, ``{AaveV2Provider contract},'' see
  \url{%https://etherscan.io/address/0x7ac6859e69d6549b39a8367097d7ae5fbff5951e#readContract}.

\bibitem{VyperContr}
------, ``{Vyper contract},'' see
  \url{https://etherscan.io/address/0xba3cfea6514cf5acddeff3167df0b7a4337751bc#code}.

\bibitem{kitzler2023disentangling}
S.~Kitzler, F.~Victor, P.~Saggese, and B.~Haslhofer, ``Disentangling
  decentralized finance (defi) compositions,'' \emph{ACM Transactions on the
  Web}, vol.~17, no.~2, pp. 1--26, 2023.

\bibitem{lee2020measurements}
X.~T. Lee, A.~Khan, S.~Sen~Gupta, Y.~H. Ong, and X.~Liu, ``Measurements,
  analyses, and insights on the entire ethereum blockchain network,'' in
  \emph{Proceedings of The Web Conference}, 2020, pp. 155--166.

\bibitem{heimbach2021behavior}
L.~Heimbach, Y.~Wang, and R.~Wattenhofer, ``Behavior of liquidity providers in
  decentralized exchanges,'' \emph{arXiv preprint arXiv:2105.13822}, 2021.

\bibitem{defillama_apis}
{DeFiLlama}, ``{APIs},'' available at: \url{https://defillama.com/docs/api}.

\bibitem{carter2021defi}
N.~Carter and L.~Jeng, ``Defi protocol risks: The paradox of defi,''
  \emph{Regtech, suptech and beyond: innovation and technology in financial
  services” riskbooks--forthcoming}, vol.~3, 2021.

\bibitem{eichengreen2023stablecoin}
B.~Eichengreen, M.~T~Nguyen, and G.~Viswanath-Natraj, ``Stablecoin devaluation
  risk,'' \emph{WBS Finance Group Research Paper}, 2023, available at SSRN:
  \url{http://dx.doi.org/10.2139/ssrn.4460515}.

\bibitem{saggese2024assessing}
P.~Saggese, E.~Segalla, M.~Sigmund, B.~Raunig, F.~Zangerl, and B.~Haslhofer,
  ``Assessing the solvency of virtual asset service providers: are current
  standards sufficient?'' \emph{Applied Economics}, pp. 1--16, 2024.

\bibitem{doerr2021defi}
J.~F. Doerr, A.~Kosse, A.~Khan, U.~Lewrick, B.~Mojon, B.~Nolens, and T.~Rice,
  ``Defi risks and the decentralisation illusion,'' \emph{BIS Quarterly
  Review}, vol.~21, 2021.

\bibitem{fracassi2024decentralized}
C.~Fracassi, M.~Khoja, and F.~Sch{\"a}r, ``Decentralized crypto governance?
  transparency and concentration in ethereum decision-making,''
  \emph{Transparency and Concentration in Ethereum Decision-Making}, 2024.

\bibitem{kitzler2024governance}
S.~Kitzler, S.~Balietti, P.~Saggese, B.~Haslhofer, and M.~Strohmaier, ``The
  governance of decentralized autonomous organizations: A study of
  contributors’ influence, networks, and shifts in voting power,'' in
  \emph{International Conference on Financial Cryptography and Data
  Security}.\hskip 1em plus 0.5em minus 0.4em\relax Springer, 2024, pp.
  313--330.

\bibitem{broido2019scale}
A.~D. Broido and A.~Clauset, ``Scale-free networks are rare,'' \emph{Nature
  communications}, vol.~10, no.~1, pp. 1--10, 2019.

\bibitem{clauset2009power}
A.~Clauset, C.~R. Shalizi, and M.~E. Newman, ``Power-law distributions in
  empirical data,'' \emph{SIAM review}, vol.~51, no.~4, pp. 661--703, 2009.

\bibitem{coingecko}
{Coingecko}, ``{Top Crypto Categories By Market Cap},'' available at:
  \url{https://www.coingecko.com/en/categories}.

\bibitem{coinmarketcap}
{Coinmarketcap}, ``{Cryptocurrency Sectors by 24h Price Change},'' available
  at: \url{https://coinmarketcap.com/cryptocurrency-category/}.

\end{thebibliography}

\newpage
\appendix

\subsection{Implementation Details}
\label{sec:implementation}

To investigate how TVL is technically computed in practice, we devised a data extraction pipeline that captures interactions of the DefiLlama Adapters GitHub repository with its environment (e.g., nodes, websites, etc.) during TVL computation.  In this Section, we provide additional implementation details on how to intercept and record such interactions.

To integrate a project into DeFiLlama, users need to develop plugins that contain the logic for computing TVL. The DeFiLlama Adapters GitHub repository provides an SDK to simplify this integration process. Once the project is successfully integrated, a dedicated folder is created within the repository, containing the protocol plugin and the necessary information to compute TVL.

To intercept the interactions, we instrument the runtime environment and execute each project's plugin at the specified Ethereum block height and DeFiLlama commit. Refer to \url{https://github.com/mswjs/interceptors} for details on how we intercept all HTTP calls made by the Node.js process.  In this way, we systematically capture all interactions (via http calls) with the environment involved in generating the TVL for a specific project. This notably encompasses all calls directed towards the Ethereum node software, as well as interactions with external hosts.

Table~\ref{tab:tokens} reports additional information on the calls directed to the Ehtereum node. Specifically, it shows the most common tokens queried in \textit{balanceOf} calls. TVL is computed mostly on wETH and wBTC (respectively $\num{10984}$ and $1415$ queries), on stablecoins (USDC, $4236$; USDT, $2197$; DAI, $2144$), governance tokens (UNI, $891$; COMP, $790$; AAVE, $777$), and staked tokens (stETH; $654$).

\begin{table}[th]
	\centering

{\footnotesize
	\begin{tabular}{l @{\hspace{1.5cm}} l @{\hspace{1.5cm}} r}
		\toprule
		Address & Symbol & Count\\
		\midrule
		0xC02aaA3... & WETH & \num{10984} \\ 0xA0b8699... & USDC & \num{4236} \\
		0xdAC17F9... & USDT & \num{2197} \\
		0x6B17547... & DAI & \num{2144} \\
		0x2260FAC... & WBTC & \num{1415} \\
		0x5149107... & LINK & \num{1002} \\
		0x1f9840a... & UNI & \num{891} \\
		0xc00e94C... & COMP & \num{790} \\
		0x7Fc6650... & AAVE & \num{777} \\
		0xD533a94... & CRV & \num{760} \\
		0x9f8F72a... & MKR & \num{715} \\
		0x6B35950... & SUSHI & \num{690} \\
		0xae7ab96... & stETH & \num{654} \\
		0x7D1AfA7... & MATIC & \num{646} \\
		0x1111111... & 1INCH & \num{616} \\
		0x408e418... & REN & \num{615} \\
		0x4Fabb14... & BUSD & \num{570} \\
		0x0F5D2fB... & MANA & \num{564} \\
		0xc944E90... & GRT & \num{556} \\
		0x0000000... & TUSD & \num{549} \\
		\bottomrule
	\end{tabular}
}
	\captionof{table}{\textbf{Most frequently called tokens in \textit{balanceOf} functions.} Wrapped ETH and BTC, stablecoins, governance and staked tokens play a primary role.}
	\label{tab:tokens}
\end{table}

\subsection{Reproducibility and reliance on off-chain data}
\label{sec:app_repr}

In this appendix, we provide additional information on the servers and errors that were detected during TVL computation.
Table~\ref{tab:errors} reports a list of the documented errors categorized by typology. The most occurring ones are related to the block height provided not being accepted, or are caused by a collection of asynchronous operations that did not complete successfully. Other errors can be reconducted to the lack of missing fields or parameters and other technical problems.
Table~\ref{tab:servers} reports a full list of the detected servers. While the one occurring most frequently is part of TheGraph infrastructure, an open-source software used to collect, process, and store data from various blockchain applications, the majority are servers from third parties. We note that we excluded the server `coins.llama.fi', as it is likely associated with DeFiLlama itself and used for internal operations.

\begin{table}[h]
	\small
	\centering
	\begin{tabular}{l @{\hspace{0.5cm}} r}
		\toprule
		Error & Count \\
		\midrule
		Block height & 53 \\ 
		Asynchronous calls failed & 17 \\
		Key required & 5 \\
		Missing field/parameter & 5 \\
		Undefined/null object & 4 \\
		Call method failed & 3 \\
		Invalid token/balance & 2 \\
		GraphQL error & 1 \\
		\bottomrule
	\end{tabular}
	\caption{\textbf{Errors occurred during TVL computation.} Most of them are related to technical issues when running the TVL-computing code.}
	\label{tab:errors}
\end{table}

\begin{table*}[h]
	\small
	\centering
	\begin{tabular}{l l @{\hspace{1cm}} l l}
\toprule
Server & Count & Server & Count \\
\midrule
api.thegraph.com & 29 & config.rampdefi.com & 1 \\
raw.githubusercontent.com & 5 & analytics.back.popsicle.finance & 1 \\
sushi-analytics.onrender.com & 3 & api.affinedefi.com & 1 \\
rpc.ankr.com & 3 & bridge.orbitchain.io & 1 \\
vault-content-api.teahouse.finance & 2 & api.angle.money & 1 \\
tvl-adapter-cache.s3.eu-central-1.amazonaws.com & 2 & api.axelarscan.io & 1 \\
api.myso.finance & 2 & api.beefy.finance & 1 \\
crucible.alchemist.wtf & 1 & api.clipper.exchange & 1 \\
data.cian.app & 1 & api.cream.finance & 1 \\
devapi.ease.org & 1 & api.daomaker.com & 1 \\
knit-admin.herokuapp.com & 1 & api.debridge.finance & 1 \\
explorer.poly.network & 1 & api.defiedge.io & 1 \\
f8wgg18t1h.execute-api.us-west-1.amazonaws.com & 1 & api.exchange.coinbase.com & 1 \\
files.insurace.io & 1 & api.flashstake.io & 1 \\
gateway-arbitrum.network.thegraph.com & 1 & api.flokifi.com & 1 \\
counterstake.org & 1 & api.goldsky.com & 1 \\
graph-node.mainnet.termfinance.io & 1 & api.hord.app & 1 \\
graph-proxy.nftx.xyz & 1 & api.hotcross.com & 1 \\
homora-api.alphafinance.io & 1 & api.mean.finance & 1 \\
messina.one & 1 & api.multibit.exchange & 1 \\
lsd-subgraph.joinstakehouse.com & 1 & api.nodes-brewlabs.info & 1 \\
bsc-dataseed1.defibit.io & 1 & api.resonate.finance & 1 \\
metabase.internal-streamflow.com & 1 & api.staking.ankr.com & 1 \\
midgard.ninerealms.com & 1 & api.studio.thegraph.com & 1 \\
moonbeam.public.blastapi.io & 1 & api.tokensfarm.com & 1 \\
partner-api.stafi.io & 1 & api.unrekt.net & 1 \\
polygon-rpc.com & 1 & api.vesper.finance & 1 \\
preserver.mytokenpocket.vip & 1 & app.everrise.com & 1 \\
stakehouse-subgraph.joinstakehouse.com & 1 & assets.nabox.io & 1 \\
static.optimism.io & 1 & backend.mochi.fi & 1 \\
token-list.solv.finance & 1 & beaconcha.in & 1 \\
universe.staderlabs.com & 1 & bsc-dataseed.binance.org & 1 \\
\bottomrule
\end{tabular}

	\caption{\textbf{External servers utilized in TVL computations.} An occurrence is counted each time a protocol interacts with a server. While some pose a smaller threat to verifiability, e.g., TheGraph, others appear to be associated with specific protocols.}
	\label{tab:servers}
\end{table*}

\subsection{Heterogeneity of on-chain interactions}
\label{sec:app_het}

We now discuss the network analysis results 
for the frequency of occurrence in absolute terms of the functions called, 
of the number of protocols calling each specific function, 
and for the number of token calls in \textit{balanceOf} functions. 
Following established methodologies~\cite{broido2019scale,clauset2009power}, we estimate the parameters $\hat{\theta} = (\hat{k}_{min},\hat{\alpha})$ and conduct a goodness-of-fit test via a bootstrapping procedure
(N = 1,000). The resulting p-value indicates if the power law is a plausible fit for the empirical data (i.e., p $\geq$ 0.1) or not. A log-likelihood ratio ($\mathcal{R}$) test is conducted to compare the power-law fit against other heavy-tailed distributions
(exponential, lognormal, and weibull).
While the bootstrap analysis shows that the hypothesis that a power-law distribution is
a good fit holds only for the distribution of the number of protocols calling each specific function, in all three cases the power law is either a better fit with respect to the other distributions, or the test is inconclusive.
An inflection point in the distribution of the number of token calls in \textit{balanceOf} functions, between the values 400 and 1,000 of the x-axis, indicates that values on the right of the elbow are overrepresented and potentially the existence of a transition region.


\begin{table*}[h]
	\centering
	\footnotesize
	\begin{tabular}{l l l @{\hspace{1.5cm}}l l l}
\toprule
Method & Count & N of protocols & Method & Count & N of protocols \\
\midrule
balanceOf & 102655 & 641 & underlying & 1613 & 43 \\
getLockedTokenAtIndex & 44022 & 1 & escrows & 1203 & 1 \\
balanceOfUnderlying & 4191 & 3 & token0 & 1040 & 94 \\
getCurrentTokens & 3436 & 10 & token1 & 1040 & 94 \\
getReserves & 3016 & 121 & tokenByIndex & 875 & 2 \\
totalSupply & 2577 & 154 & balance & 874 & 9 \\
token & 2456 & 24 & get\_coins & 814 & 2 \\
symbol & 1835 & 64 & pool\_list & 809 & 1 \\
eth\_getBalance & 1696 & 170 & getEthBalance & 791 & 7 \\
poolInfo & 1672 & 25 & calcTotalValue & 773 & 1 \\
\bottomrule
\end{tabular}

	\caption{\textbf{Most called functions in absolute terms.} A number of functions (e.g., \textit{balanceOfUnderlying}, \textit{balance}) have names indicating that they are likely used to compute balance in a non-standard way.
	}
	\label{tab:calls_count}
\end{table*}

\begin{table*}[h]
	\centering
	\footnotesize
	\begin{tabular}{l l @{\hspace{0.5cm}}l l}
	\toprule
	Function name & Hex Signature & Function name & Hex Signature \\
	\midrule
	accountedBalance & ['0x0937eb54'] & getTotalRPLStake & ['0x9a206c8e'] \\
	allBalances & ['0x555b6162'] & getTotalReserves & ['0x242693d3'] \\
	balance & ['0xb69ef8a8'] & getTotalUnderlying & ['0xb40494e5'] \\
	balanceOf & \makecell[l]{['0x00fdd58e', '0x35ee5f87',\\'0x3656eec2', '0xf7888aec']} & getTotalValueInPool & ['0xc8ecaf30'] \\
	balanceOfUnderlying & ['0x3af9e669'] & getTrackedAssets & ['0xc4b97370'] \\
	balances & \makecell[l]{['0x4903b0d1', '0x065a80d8',\\'0x8909aa3f', '0x27e235e3']} & getTvl & ['0xd075dd42'] \\
	borrowBalanceStored & ['0x95dd9193'] & getUnderlying & ['0x9816f473'] \\
	calcTotalValue & ['0xc7de38a6'] & getUnderlyingBalances & ['0x1322d954'] \\
	checkBalance & ['0x5f515226'] & getUnderlyings & ['0xf65baefa'] \\
	currencyBalance & ['0x5c75347a'] & \makecell[l]{getUnderlyingsAmounts-\\FromClusterAmount} & ['0x9bb1bebb'] \\
	currentTotalStake & ['0xce4843e9'] & lockedBalances & ['0x0483a7f6'] \\
	getAllAssets & ['0x2acada4d'] & lockedLiquidityOf & ['0xd9f96e8d'] \\
	getAllStakes & ['0x04238994'] & lockedStakesOf & ['0x1e090f01'] \\
	getAssets & ['0x67e4ac2c'] & lockedSupply & ['0xca5c7b91'] \\
	getBassets & ['0x1d3ce398'] & poolBalance & ['0x96365d44'] \\
	getCacheBalances & ['0x4a9d1036'] & syncBalance & ['0xfd9c652b'] \\
	getContractValue & ['0xdc82697c'] & totalAsset & ['0xf9557ccb'] \\
	getETHPx & ['0xab9aadfe'] & totalAssetAmount & ['0xfd27152c'] \\
	getEthBalance & ['0x4d2301cc'] & totalAssetBorrow & ['0x20f6d07c'] \\
	getLockedVestings & ['0x344e58d3'] & totalAssets & ['0x01e1d114'] \\
	getPoolAmount & ['0x945eb764'] & totalBalance & ['0xad7a672f'] \\
	getPoolTotalValue & ['0xf1437c16'] & totalBalanceOf & ['0x4b0ee02a'] \\
	getRawFundBalances & ['0xe2e4c60c'] & totalETH & ['0x36bdee74'] \\
	\makecell[l]{getRawFund-\\BalancesAndPrices} & ['0x0d8f8a90'] & totalReserve & ['0x4c68df67'] \\
	getReserveTotalBorrows & ['0xe6d18190'] & totalReserves & ['0x8f840ddd'] \\
	getSupply & ['0xf77ee79d'] & totalSYNCLocked & ['0xc3f4d79f'] \\
	getSupportedAsset & ['0x60a8b18a'] & totalStaked & ['0x817b1cd2'] \\
	getSupportedAssets & ['0xe5406dbf'] & totalTokenBalanceStakers & ['0x1878fbf3'] \\
	getSupportedAssetsLength & ['0xc0fd22b7'] & totalValue & ['0xd4c3eea0'] \\
	getToken1Balance & ['0x5153786b'] & total\_staked & ['0xaf7568dd'] \\
	getTotalAmounts & ['0xc4a7761e'] & underlyingBalance & ['0x59356c5c'] \\
	getTotalAsset & ['0x2768385d'] & virtualBalance & ['0xdcd2af17'] \\
	getTotalBalance & ['0x12b58349'] & virtualUsdtAccumulatedBalance & ['0xd88953b4'] \\
	getTotalPooledEther & ['0x37cfdaca'] & yvCurveFRAXBalance & ['0xc645065e'] \\
	\bottomrule
\end{tabular}

	\caption{\textbf{Alternative functions to balanceOf likely used to compute TVL.} For each function we report the name ($1^{st} column)$ and its hex signature ($2^{nd} column)$.}
	\label{tab:alt_fcts}
\end{table*}


Next, we comment the approach used to identify alternative functions likely used to compute TVL. 
Table~\ref{tab:calls_count} reports the most called on-chain functions in absolute terms, rather than being ranked by the number of protocols that call them. We notice that a number of function names (\textit{balanceOfUnderlying}, \textit{balance}, as well as \textit{totalAssets} that appeared in Table~\ref{tab:calls}), indicate functions that are likely to compute balance in alternative ways with respect to the \textit{balanceOf} most common method. %
To identify all these functions, we use a set of regular expressions that capture any method whose function name includes the term `balance', or a case-insensitive combination of the following terms: `total, get, locked' AND `tvl, Ether, ETH, stake, underlying, reserve, amount, supply, value, locked, shares, asset, liquidity'. Furthermore, we conduct a manual check to remove functions that are clearly not computing TVL but provide supplementary functionalities, such as \textit{totalSupply} or \textit{getReserves}. 
Table~\ref{tab:alt_fcts} reports the full list of alternative functions to \textit{balanceOf} likely used to compute TVL.
In total, we remove the following 21 functions: \textit{getAssetInfo}, \textit{getAssetsPrices}, \textit{getAssetsWithState},  \textit{getBNFTAssetList}, 
\textit{getLiquidityPools}, \textit{getLockedTokenAtIndex}, \textit{getNumLockedTokens}, \textit{getReserveData}, \textit{getReserves}, \textit{getReservesData}, 
\textit{getReservesList},  \textit{getUnderlyingAsset}, \textit{getUnderlyingOfIBTAddress}, \textit{getUnderlyingPrice}, \textit{getUnderlyingTokenAddress},
\textit{totalShares}, \textit{totalSupply}, \textit{totalUnderlying}, \textit{totalUnderlyingSupply}.

\subsection{Changes to computation methods over time}
\label{sec:app_changes}

The analysis in Section~\ref{sec:commits_comp} follows the intuition that developers can modify how TVL is computed and that the collection may be dependent on one specific implementation, thus TVL computations might change across time and commits. We thus repeat the data gathering on the following commits:

\begin{itemize}
	\item \small{6764756f9270ab6a3047c06c13c0b1b2d32a3247: Jan 4 2024;}
	\item \small{aef637c6413b5101667e567a0422769b1ca99564: Oct 4 2023;}
	\item \small{1bfb7b5c798c7a491694882a7b390ed41385c315: Jul 4 2023;}
	\item \small{72b764c5d0bb5896f2857dd8c2ced5e89e7fb063: Apr 4 2023;}
	\item \small{679f52e123a19d9c5160d1aa79a33dd9de6dc5ec: Jan 4 2023;}
\end{itemize}
As discussed in the main body of the paper, we do not analyze data earlier than 2023: the structure of some critical files within the DeFiLlama repository has changed in time, thus making our data-gathering infrastructure less reliable on older commits. In particular, the number of comparable commits decreases steadily with time, and for commits older than Aug 2021, our pipeline becomes not compatible with specific changes made to the repository (see \url{https://github.com/DefiLlama/DefiLlama-Adapters/discussions/432}).
Table~\ref{tab:commits_stats} shows that differences in the dataset of captured calls across commits exist: the total amount of calls to an Ethereum node is not stable over time, and few specific protocols play a relevant role in this sense: as one can see in Table~\ref{tab:commits_stats}, the differences are markedly smaller when some specific projects are excluded.

\begin{table}[ht]
	\footnotesize
	\centering
\begin{tabular}{l @{\hspace{0.5cm}} r @{\hspace{0.3cm}} r  @{\hspace{0.3cm}} r  @{\hspace{0.3cm}} r  @{\hspace{0.3cm}} r}
\toprule
 & Jan 24 & Oct 23 & Jul 23 & Apr 23 & Jan 23 \\
\midrule
Total calls & \num{197775} & \num{333239} & \num{262512} & \num{314280} & \num{254259} \\
Unicrypt calls & \num{44032} & \num{198228} & \num{150094} & \num{104165} & \num{77063} \\
Other calls & \num{153743} & \num{135011} & \num{112418} & \num{210115} & \num{177196} \\
\bottomrule
\end{tabular}

	\caption{\textbf{Dataset dimension in different commits}. One specific protocol (Unicrypt) is responsible for large variations across commits.}
	\label{tab:commits_stats}
\end{table}

\begin{figure}[h!]
	\centering
	\begin{subfigure}[b]{0.4\textwidth}
		\centering
		\includegraphics[width=\textwidth]{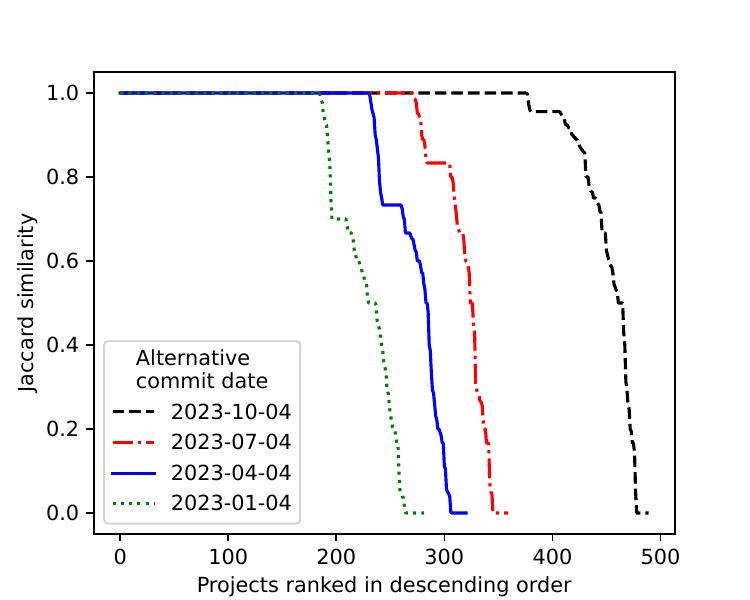}
		\caption{Figure~\ref{fig:comp_main}, x-axis non normalized}
		\label{fig:comp_non_norm}
	\end{subfigure}
	\hfill
	\begin{subfigure}[b]{0.4\textwidth}
		\centering
		\includegraphics[width=\textwidth]{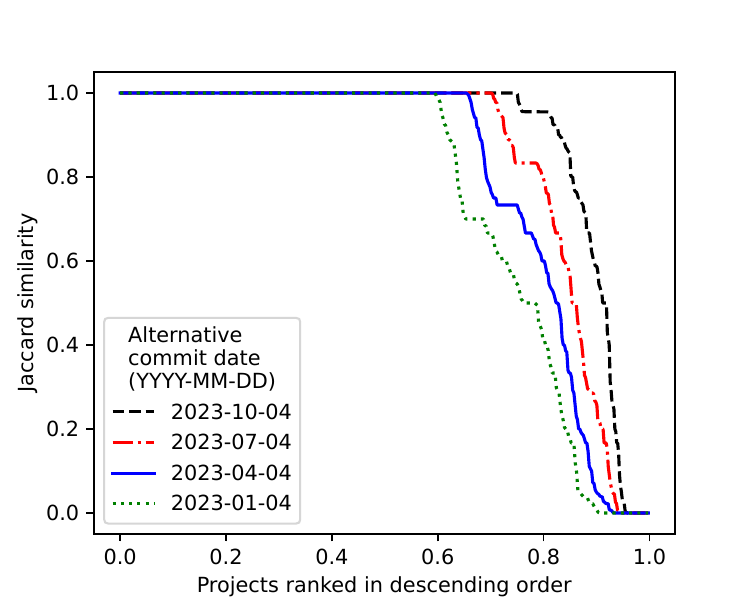}
		\caption{Figure~\ref{fig:comp_main}, all protocols}
		\label{fig:comp_with_errs}
	\end{subfigure}
	\caption{\textbf{Jaccard similarity.} Panel~(\subref{fig:comp_non_norm}) reports the same information of Figure~\ref{fig:comp_main}, but projects are reported on the x-axis in absolute terms instead of being normalized. The sample of comparable protocols decreases sharply when older commits are compared. Panel~(\subref{fig:comp_with_errs}) shows instead the Jaccard similarity values when including also projects that raised errors in the data-gathering process or used external hosts. Average values range from 0.90\% to 0.75\%.}
	\label{fig:other_comps}
\end{figure}

Next, in Figure~\ref{fig:other_comps} we report additional information on the Jaccard similarity analysis. 
We recall that the Jaccard similarity coefficient, utilized for measuring the similarity of overlapping sets, is defined as the ratio between their intersection and their union, and ranges between 0 and 1. 
We therefore favor it over alternative similarity metrics that measure distances between text strings (e.g., Hamming), vectors (e.g., cosine similarity) or that identify correlations across data points (e.g., Spearman or Pearson). The left panel shows the same information of Figure~\ref{fig:comp_main}, but the x-axis is not normalized. The right panel includes instead all protocols, therefore also the ones that raised errors in the data-gathering process or used external hosts. The average Jaccard values are 0.90\%, 0.85\%, 0.80\%, and 0.75\% respectively for Oct, Jul, Apr, and Jan 2023.
Notably, we cannot exclude that differences in older commits are due to a less reliable collection. Therefore, our findings can be interpreted as a lower boundary for the similarity values.

Finally, Table~\ref{tab:commit_data} reports information on the evolution over time of the ratio between the number of standard and alternative balance queries (excluding protocols that used hosts or raised errors during the data collection).

\begin{table}[h!]
	\centering
	\begin{tabular}{l @{\hspace{0.5cm}} r @{\hspace{0.3cm}} r  @{\hspace{0.3cm}} r  @{\hspace{0.3cm}} r  @{\hspace{0.3cm}} r}
		\toprule
		& \multicolumn{5}{c}{Commit identifiers and date} \\
		\cmidrule{2-6}
		& \makecell{676475\\(2024 \\01-04)} & \makecell{aef637\\(2023 \\10-04)} & \makecell{1bfb7b\\(2023\\ 07-04)} & \makecell{72b764\\(2023 \\04-04)} & \makecell{679f52\\(2023\\ 01-04)} \\
		\midrule
		Alt. functions & 6831 & 7554 & 6836 & 6882 & 7197 \\
		Std. balance queries      & 70073 & 54144 & 27534 & 23426 & 18367 \\
		Ratio    & 0.089 & 0.122 & 0.199 & 0.227 & 0.282 \\
		\bottomrule
	\end{tabular}
	\caption{\textbf{Evolution over time of the ratio between the number of standard and alternative balance queries.} Standard balance queries include \textit{balanceOf} and \textit{eth\_getBalance} calls.}
	\label{tab:commit_data}
\end{table}

\subsection{Non-mutually exclusive smart contract calls}
\label{sec:app_dc}

We report the list of duplicated addresses in Tables~\ref{tab:dc_eth} and~\ref{tab:dc_bof}. These correspond to calls executed by different protocols (column `Protocols') on the same input address (column `Input') and token address (only for Table~\ref{tab:dc_bof}, column `On'). Upon closer inspection, we find that in most cases they are related to interconnected protocols (see, e.g., metis and metisBridge and Curve and Bent). One possible explanation for this phenomenon is that these addresses are reported directly by the project developers; we cannot exclude that some of the smart contracts are incorrectly reported multiple times only for a temporary time span and that they are further removed by the maintainers of the DeFiLlama platform.

\begin{table}[h]
	\hspace*{-0.2cm}
	\centering
	\footnotesize
	\begin{tabular}{ll}
\toprule
Input & Protocols \\
\midrule
0x3980c9ed79d2c191a89e02fa3529c60ed6e9c04b & ['metis' 'metisBridge'] \\
0x8301ae4fc9c624d1d396cbdaa1ed877821d7c511 & ['curve' 'bent'] \\
0xb576491f1e6e5e62f1d8f26062ee822b40b0e0d4 & ['curve' 'bent'] \\
0xd51a44d3fae010294c616388b506acda1bfaae46 & ['curve' 'bent'] \\
0xdc24316b9ae028f1497c275eb9192a3ea0f67022 & ['curve' 'bent'] \\
\bottomrule
\end{tabular}

	\caption{\textbf{Duplicated get\_ethBalance functions}. get\_ethBalance calls executed by different protocols on the same input address (column `Input').}
	\label{tab:dc_eth}
\end{table}

\begin{table*}[h!]
	\centering
	\scriptsize
	\begin{tabular}{llllll}
\toprule
Input & On & Protocols & Input & On & Protocols \\
\midrule
0x031816fd... & 0x03e173ad... & ['dodo' 'thales'] & 0x031816fd... & 0xc02aaa39... & ['dodo' 'thales'] \\
0x0f41eade... & 0xc36442b4... & ['pawnfi-lending' 'pawnfi-nft'] & 0x16770d64... & 0xc02aaa39... & ['enzyme' 'diva'] \\
0x19b080fe... & 0x96e61422... & ['keep3r' 'curve'] & 0x1a26ef65... & 0x42bbfa2e... & ['bobagateway' 'boba'] \\
0x1a26ef65... & 0xd26114cd... & ['bobagateway' 'boba'] & 0x1ce8aafb... & 0xae7ab965... & ['enzyme' 'diva'] \\
0x23012599... & 0xba30e5f9... & ['pawnfi-lending' 'pawnfi-nft'] & 0x25d6fe0d... & 0x49cf6f5d... & ['pawnfi-lending' 'pawnfi-nft'] \\
0x27e49962... & 0x790b2cf2... & ['pawnfi-lending' 'pawnfi-nft'] & 0x27f23c71... & 0xc02aaa39... & ['enzyme' 'nexus'] \\
0x306b1950... & 0xeca82185... & ['uma' 'perlinx'] & 0x325a0e5c... & 0xae7ab965... & ['swell-vault' 'enzyme'] \\
0x325a0e5c... & 0xc02aaa39... & ['swell-vault' 'enzyme'] & 0x32ecc1de... & 0xb7f7f6c5... & ['pawnfi-lending' 'pawnfi-nft'] \\
0x3980c9ed... & 0x1f9840a8... & ['metis' 'metisBridge'] & 0x3980c9ed... & 0x2260fac5... & ['metis' 'metisBridge'] \\
0x3980c9ed... & 0x3405a1bd... & ['metis' 'metisBridge'] & 0x3980c9ed... & 0x4fabb145... & ['metis' 'metisBridge'] \\
0x3980c9ed... & 0x51491077... & ['metis' 'metisBridge'] & 0x3980c9ed... & 0x6226e00b... & ['metis' 'metisBridge'] \\
0x3980c9ed... & 0x6b175474... & ['metis' 'metisBridge'] & 0x3980c9ed... & 0x6b359506... & ['metis' 'metisBridge'] \\
0x3980c9ed... & 0x7fc66500... & ['metis' 'metisBridge'] & 0x3980c9ed... & 0x9e32b13c... & ['metis' 'metisBridge'] \\
0x3980c9ed... & 0xa0b86991... & ['metis' 'metisBridge'] & 0x3980c9ed... & 0xba6b0dbb... & ['metis' 'metisBridge'] \\
0x3980c9ed... & 0xd533a949... & ['metis' 'metisBridge'] & 0x3980c9ed... & 0xdac17f95... & ['metis' 'metisBridge'] \\
0x3a93e863... & 0xeca82185... & ['uma' 'perlinx'] & 0x3e75dcad... & 0xa0b86991... & ['domfi' 'uma'] \\
0x3f1b0278... & 0xfafdf0c4... & ['keep3r' 'curve'] & 0x41284a88... & 0x0bc529c0... & ['percent' 'balancer-v1'] \\
0x41284a88... & 0xc02aaa39... & ['percent' 'balancer-v1'] & 0x43b4fdfd... & 0xbc6da0fe... & ['curve' 'bent'] \\
0x46f5e363... & 0xeca82185... & ['uma' 'perlinx'] & 0x4a2f0ca5... & 0x1f573d6f... & ['bancor' 'ichifarm'] \\
0x4a2f0ca5... & 0x903bef17... & ['bancor' 'ichifarm'] & 0x4e8d60a7... & 0xc02aaa39... & ['degenerative' 'uma'] \\
0x4f1424ce... & 0xa0b86991... & ['degenerative' 'uma'] & 0x505efcc1... & 0x04abeda2... & ['nest' 'parasset'] \\
0x516f5959... & 0xc02aaa39... & ['degenerative' 'uma'] & 0x55a8a39b... & 0x99d8a9c4... & ['curve' 'bent'] \\
0x55a8a39b... & 0xa47c8bf3... & ['curve' 'bent'] & 0x58378f5f... & 0x903bef17... & ['ichifarm' 'balancer-v1'] \\
0x58378f5f... & 0xc02aaa39... & ['ichifarm' 'balancer-v1'] & 0x5a6a4d54... & 0x99d8a9c4... & ['curve' 'bent'] \\
0x5eeaef7d... & 0xed5af388... & ['pawnfi-lending' 'pawnfi-nft'] & 0x5f0a4a59... & 0xbc4ca0ed... & ['pawnfi-lending' 'pawnfi-nft'] \\
0x7514799c... & 0xe012baf8... & ['pawnfi-lending' 'pawnfi-nft'] & 0x799c9518... & 0xa0b86991... & ['degenerative' 'uma'] \\
0x7c62e5c3... & 0xc02aaa39... & ['degenerative' 'uma'] & 0x7d0b6fb1... & 0x60e4d786... & ['pawnfi-lending' 'pawnfi-nft'] \\
0x82c427ad... & 0xc02aaa39... & ['opyn-squeeth' 'uniswap'] & 0x8301ae4f... & 0xc02aaa39... & ['curve' 'bent'] \\
0x8301ae4f... & 0xd533a949... & ['curve' 'bent'] & 0x8461a004... & 0x95dfdc81... & ['keep3r' 'curve'] \\
0x8818a9bb... & 0x5555f75e... & ['keep3r' 'curve'] & 0x94e653af... & 0xa0b86991... & ['domfi' 'uma'] \\
0x99e58237... & 0x6b175474... & ['percent' 'balancer-v1'] & 0x99e58237... & 0xc02aaa39... & ['percent' 'balancer-v1'] \\
0x9a5c88ac... & 0x04abeda2... & ['nest' 'parasset'] & 0x9c2c8910... & 0x1cc481ce... & ['keep3r' 'curve'] \\
0x9d046499... & 0x62b9c735... & ['curve' 'bent'] & 0x9d046499... & 0xd533a949... & ['curve' 'bent'] \\
0x9fe9bb6b... & 0x9ea3b5b4... & ['delta' 'core'] & 0xaa5a67c2... & 0x86537736... & ['curve' 'bent'] \\
0xaff95ac1... & 0x903bef17... & ['rari' 'ichifarm'] & 0xb1a3e5a8... & 0xa0b86991... & ['degenerative' 'uma'] \\
0xb39edbc5... & 0xe0a97733... & ['tokensfarm' 'bloxmove'] & 0xb40ba947... & 0xeca82185... & ['uma' 'perlinx'] \\
0xb576491f... & 0x4e3fbd56... & ['curve' 'bent'] & 0xb576491f... & 0xc02aaa39... & ['curve' 'bent'] \\
0xba3436fd... & 0x1a7e4e63... & ['angle' 'curve'] & 0xba3436fd... & 0x1abaea1f... & ['angle' 'curve'] \\
0xbaaa1f5d... & 0x853d955a... & ['curve' 'bent'] & 0xbaaa1f5d... & 0x956f47f5... & ['curve' 'bent'] \\
0xbaaa1f5d... & 0xbc6da0fe... & ['curve' 'bent'] & 0xbebc4478... & 0x6b175474... & ['curve' 'bent'] \\
0xbebc4478... & 0xa0b86991... & ['curve' 'bent'] & 0xbebc4478... & 0xdac17f95... & ['curve' 'bent'] \\
0xc3160c5c... & 0x40803cea... & ['spool-v2' 'spool'] & 0xc48b8329... & 0xc00e94cb... & ['percent' 'balancer-v1'] \\
0xc48b8329... & 0xc02aaa39... & ['percent' 'balancer-v1'] & 0xc697051d... & 0x7fc66500... & ['aave' 'balancer-v1'] \\
0xc697051d... & 0xc02aaa39... & ['aave' 'balancer-v1'] & 0xca2531b9... & 0xc02aaa39... & ['degenerative' 'uma'] \\
0xceaf7747... & 0xa693b19d... & ['curve' 'bent'] & 0xd3a0e00f... & 0xa0b86991... & ['domfi' 'uma'] \\
0xd50fbace... & 0xeca82185... & ['uma' 'perlinx'] & 0xd51a44d3... & 0x2260fac5... & ['curve' 'bent'] \\
0xd51a44d3... & 0xc02aaa39... & ['curve' 'bent'] & 0xd51a44d3... & 0xdac17f95... & ['curve' 'bent'] \\
0xd632f226... & 0x6c3f90f0... & ['fraxfinance' 'bent'] & 0xd632f226... & 0x853d955a... & ['curve' 'bent'] \\
0xd6ac1cb9... & 0x69681f8f... & ['keep3r' 'curve'] & 0xdc24316b... & 0xae7ab965... & ['curve' 'bent'] \\
0xdcef968d... & 0xa0b86991... & ['curve' 'fraxfinance'] & 0xe010fcda... & 0x51491077... & ['percent' 'balancer-v1'] \\
0xe010fcda... & 0xc02aaa39... & ['percent' 'balancer-v1'] & 0xe867be95... & 0xba100000... & ['percent' 'balancer-v1'] \\
0xe867be95... & 0xc02aaa39... & ['percent' 'balancer-v1'] & 0xe969991c... & 0xa0b86991... & ['percent' 'balancer-v1'] \\
0xe969991c... & 0xc02aaa39... & ['percent' 'balancer-v1'] & 0xeb85b2e1... & 0xbc16da9d... & ['percent' 'balancer-v1'] \\
0xeb85b2e1... & 0xc02aaa39... & ['percent' 'balancer-v1'] & 0xee9a6009... & 0x2260fac5... & ['percent' 'balancer-v1'] \\
0xee9a6009... & 0xc02aaa39... & ['percent' 'balancer-v1'] & 0xf083fba9... & 0x4e3fbd56... & ['curve' 'bent'] \\
0xf083fba9... & 0x9e0441e0... & ['curve' 'bent'] & 0xf35a80e4... & 0xc02aaa39... & ['degenerative' 'uma'] \\
0xf861483f... & 0x853d955a... & ['fpi' 'curve'] & 0xf8ef02c1... & 0xc02aaa39... & ['degenerative' 'uma'] \\
0xfcfc434e... & 0x1498bd57... & ['delta' 'core'] & NaN & NaN & NaN \\
\bottomrule
\end{tabular}

	\caption{\textbf{Duplicated balanceOf functions}. BalanceOf calls executed by different protocols on the same token address (column `On') and input address (column `Input').}
	\label{tab:dc_bof}
\end{table*}

\subsection{TVL reconstruction from on-chain data}
\label{sec:app_reconstr}

To conduct the analyses in Section~\ref{sec:app_reconstr}, we collect on-chain data for each protocol-specific address and its associated tokens as follows.
We inspect calls that represent balance queries for Ether or tokens, assuming that addresses queried during computation contribute to the project TVL. We obtain a list of addresses contributing towards the locked value for each project, along with a compilation of tokens ($N = 12246$) in which all projects hold value.
Next, for each protocol, we extract cryptoassets balances and prices. 
To obtain quantities, we use the lists of protocol-specific addresses and tokens to query their state and extract historical balance information monthly from January $1^{st}$ 2021 to February  $1^{st}$ 2024. To price tokens in a common currency, for each token we search for trading pairs between that token and wrapped Ether (wETH) on the Uniswap V2 DEX until January 2024, extract the logged liquidity changes occurring after any trading or liquidity provision action, and compute their implied exchange rate determined by the liquidity ratio as in~\cite{heimbach2021behavior}. 
We exclude tokens with very low liquidity and remove price outliers (more details below); in total, using this approach, we could price 942 tokens. For comparisons with USD, we gather historical data from Coingecko on the ETH/USD exchange rate.

Finally, we categorize tokens following the conceptualization of Section~\ref{sec:background}. We extract data from Coingecko and Coinmarketcap~\cite{coingecko,coinmarketcap} to identify stablecoins and governance tokens, and complement this with a semi-automated search of derivative tokens and additional governance tokens based on regular expressions on their names and symbols. We divide tokens into seven different categories: Ether and its wrapped token wETH, wrapped BTC (wBTC), non-crypto-backed stablecoins, crypto-backed stablecoins, governance tokens, derivative tokens, and others (non categorized) tokens.
Specifically, we conduct the following steps:

\begin{itemize}
	\item \textbf{Stablecoins}: we extract information from Coingecko \cite{coingecko} and Coinmarketcap~\cite{coinmarketcap} on N = 64 stablecoins. We select those with more than 20 mln\$ market capitalization, whose price is stable over time, and pegged to the US dollar (stablecoins pegged to EUR or commodities like gold clearly represent a minor market). Next, we look into their documentation and distinguish those that are decentralized and backed by other cryptoassets\footnote{DAI, Liquity USD, FRAX, Ethena USDe, Decentralized USD, mkUSD, sUSD, Alchemix USD, DOLA, MIM.} (N = 11) and those that are issued by trusted entities after a deposit of external assets such as cash, treasury bills, or other cash equivalents\footnote{USDT, USDC, True USD, BUSD, PAX Dollar, First Digital USD, Paypal USD, Gemini Dollar, Verified USD.} (N = 9).
	\item \textbf{Governance tokens}: we exploit again Coingecko and Coinmarketcap to extract a list of N = 110 governance tokens; we complement it with a regular-expression-based search on the dataset of \num{12246} tokens being called in DeFiLlama to compute protocols balances and label additional tokens that contain the word `governance' in their name, for a total of N = 162 tokens;
	\item \textbf{Bitcoin tokenized representations through bridges}: whilst other tokens in addition to wrapped Bitcoin (wBTC) exist, wBTC is by far the largest one in the ecosystem. For instance, renBTC and Houobi BTC, two of the main alternatives to wBTC, have a market capitalization of around 20 mln\$ against almost 9 bnl\$ of wBTC at the time of writing. Following a similar line of thought, we only focus on Bitcoin, and do not investigate other distributed ledgers and their native tokens. 
	\item \textbf{Derivative tokens}: we utilize the dataset of \num{12246} tokens called in DeFiLlama to identify them. Upon a deep inspection of the data, we found that token names follow specific patterns; for instance, DEX LP token names typically contain terms like `LP' or `Pool'; interest-bearing tokens from protocols like Compound and Aave produce cTokens and aTokens that represent a claim on the underlying asset; similarly, sTokens are receipt tokens for the Synthetix protocol. We devise a procedure based on regular expression searches executed on the token names. We identify a total of \num{2222} derivative tokens.
\end{itemize}

We now provide additional details on the procedure envisaged to obtain token prices. As the price time series extracted are noisy, we exclude the tokens having less than \num{10000} points, i.e. those that are traded rarely and have very low liquidity, assuming that they do not play a relevant role in the ecosystem. 
Next, we implement a cleaning procedure to remove from each price time series the outliers, i.e., individual price values that diverge by several orders of magnitude from the trend. To remove them, we compute the centered rolling average on a time window of ten days and remove all values where the difference between the rolling average and the price time series is larger than 20\%. While the approach is relatively simple, it provides satisfying results for most tokens, but it is not effective on tokens with low liquidity. 
We then conduct manual sanity checks to ensure that prices are cleaned correctly and additionally remove the prices of 80 tokens whose values have exceedingly high volatility and therefore it was not possible to determine accurately their correct trend.
Future work could investigate these choices more thoroughly and provide an alternative scenario that includes also the removed tokens. However, as for most protocols the value is concentrated in few and widely used tokens, we do not expect findings to change substantially.
Finally, we manually check that the stablecoins included in the study did not deviate from their peg and assume that their price is anchored to the US dollar.

\begin{figure}[htbp]
	\centering
	\includegraphics[width=\linewidth]{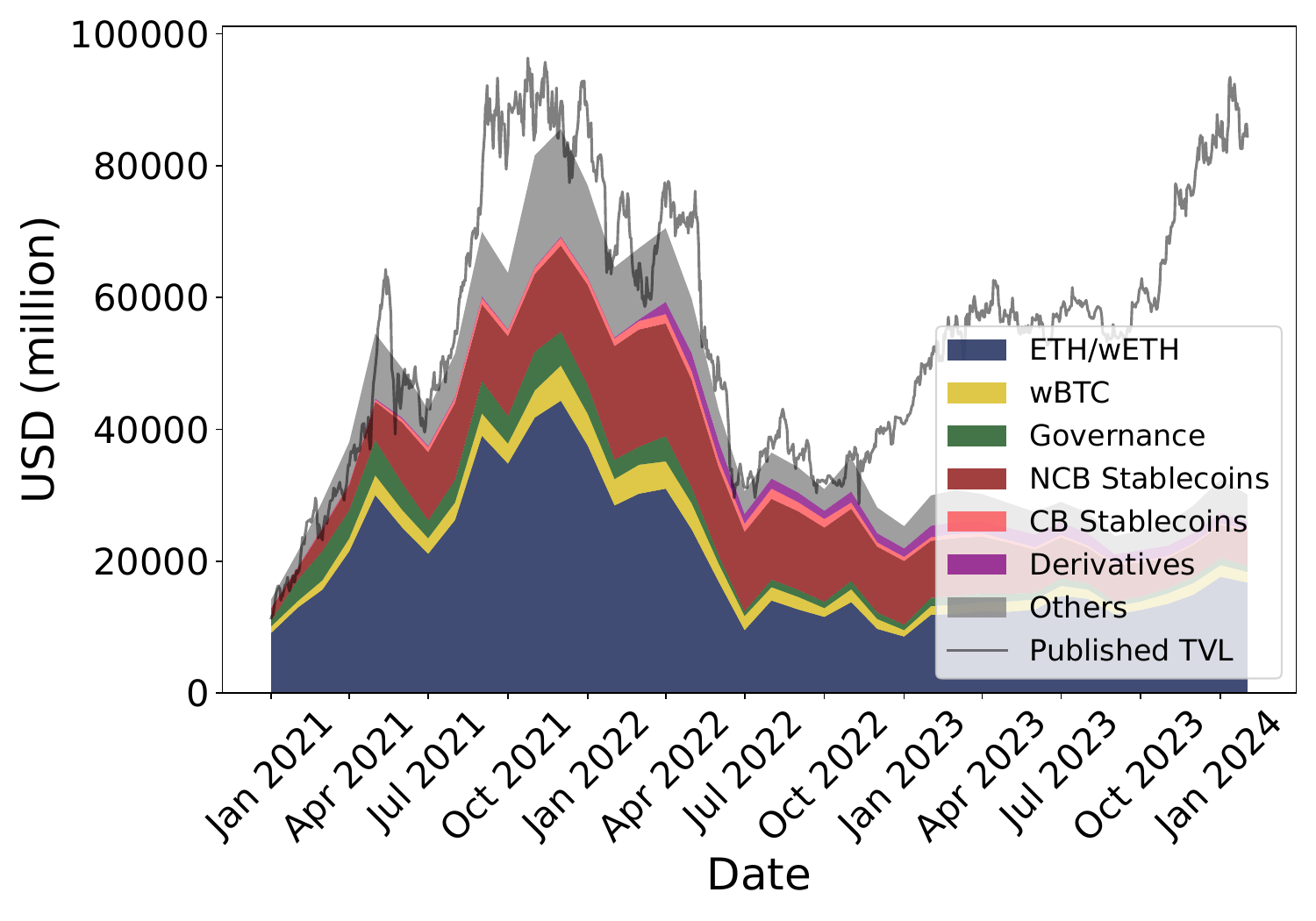}
	\caption{\textbf{TVL reconstruction for the Ethereum DeFi ecosystem.} The plot shows the evolution in time of the vTVL value for all protocols in the case study.}
	\label{fig:all_stack}
\end{figure}

To provide a broader picture of the ecosystem, in addition to the individual protocols described in Section~\ref{sec:res_tvl}, we show in Figure~\ref{fig:all_stack} the sum of the vTVL of all protocols for which we could recompute values and its evolution in time, again compared to DeFiLlama API data. 
The off-chain TVL value reported here for comparison is smaller than the TVL of the entire DeFi ecosystem, first because it is computed only on Ethereum, and second because our dataset includes only projects for which we could extract both on-chain and off-chain data.
We observe that the vTVL is dominated by Ether and non-crypto-backed stablecoins; the remaining token categories play a less relevant role. 
It is also possible to notice that the off-chain and on-chain sources correspond only until the end of 2022. 
However, we argue that in this context the comparison between on-chain data and DeFiLlama API data is for reference and illustrative purpose only.
Indeed, we noticed that for some protocols the API data are reported only after a certain date, while we find on-chain assets associated with them also before that date. This inflates the quantity of on-chain assets in comparison with the off-chain ones for earlier dates.
Second, our infrastructure did not capture standard balance calls enabling to reconstruct the TVL of the protocol Lido, which is one of the largest protocols in terms of TVL according to DeFiLlama, especially after the end of 2022. 
Third, by investigating the largest protocols, we found that the API files do not always report data for protocol versions separated consistently with respect to how they are reported in the GitHub DeFillama repositories. For instance, Aave APIs report together Aave v2 and v3 (the repository reports one folder for Aave-v1 and one named `Aave' seemingly for v2 and v3). The Uniswap API file reports together v2 and v3; however, Uniswap-v1, -v2, and -v3 are reported in a separate folder (we thus corrected API data accordingly). We also note that Uniswap -v1 and -v2 values are partly computed through the use of external hosts, while for Uniswap v3 we captured on-chain interactions. 
Similarly, a steadily discrepancy emerges for the protocol Maker at the end of 2022. Also in this case, as explained in the main body of the paper, this could be due to the use of external hosts to compute TVL, but this pattern is also consistent with the possibility that the APIs include both Maker and its related project Maker RWA.
All these considerations explains at least in part the structural difference that emerges in the comparison of on-chain and off-chain sources between and after end of 2022. Finally, as the change also corresponds with the Merge (which took place on September 2022), we cannot exclude that also this event has a role. This aspect needs further investigation.

\subsection{TVL Composition Changes Across Categories \& Time}
\label{sec:app_tvr}

In this section we examine TVL composition changes by posing specific attention to the tokens included in the calculation of TVR introduced by Luo et al.~\cite{luo2024piercing}.
Specifically, we analyze how the ratio between the value of assets used to measure TVR and the value of TVL, $ R = \frac{TVR_a}{TVL} $, changes across protocol category, protocol size, and time. 

\begin{figure}[htbp]
	\centering
	\includegraphics[width=\linewidth]{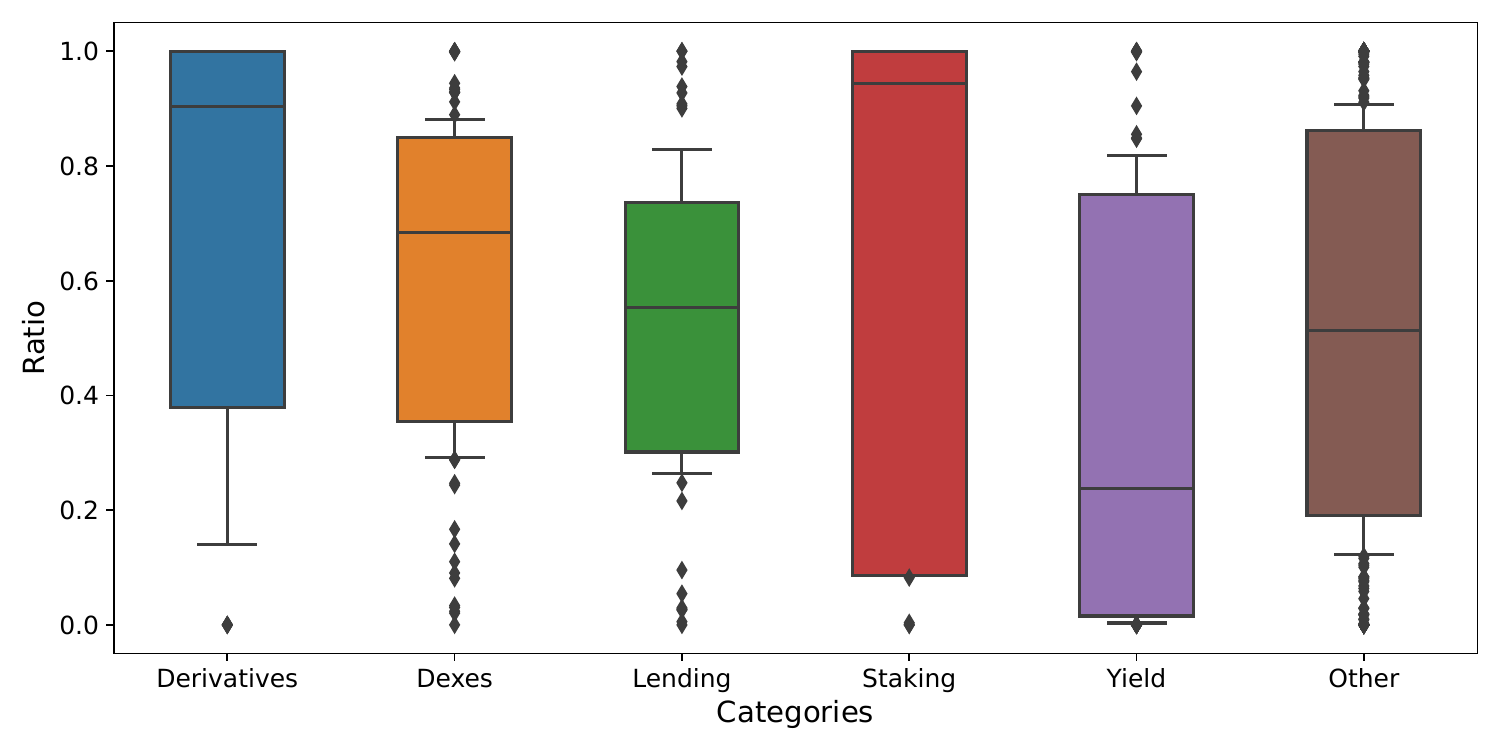}
	\caption{\textbf{Ratio R for different protocol categories}. We compute $ R = TVR_a/TVL $ as the ratio between the value of assets used to measure TVR and the value of TVL, and compare it across protocol categories. Derivtives and staking protocols have the highest R values.}
	\label{fig:boxplot}
\end{figure}

We utilize the API DeFiLlama data to avoid data loss and utilize our token categorization to compute the amount of each protocol as the US dollar amount of plain tokens, i.e., native tokens, governance tokens, and NCB stablecoins. 
We focus on the subset of protocols (N = 412) with average TVL larger than $5 \cdot 10^{4}$ USD to ensure noise reduction and group them in DeFi categories as reported by DeFiLlama, focusing on Derivatives (N = 19), DEXs (N = 69), Lending (N = 46), Staking (N = 17), and Yield (N = 47) protocols. The remaining ones (including Collateralized Debt Position or CDP protocols, Services, Real World Assets or RWA, NFT Marketplaces, \dots) are categorized as `Others'.
Figure~\ref{fig:boxplot} reports the ratio R for these protocol categories without distinguishing them based on size or time. We find that the median value of R is respectively 90.4\%, 68.3\%, 55.3\%, 94.4\%, 23.8\%, and 51.4\%.
Yield protocols highly rely on non-redeemable tokens, while Derivatives and Staking protocols have the highest ratio, indicating lower reliance on them.

\begin{figure}[t]
	\centering
	\begin{subfigure}[b]{\linewidth}
		\centering
		\includegraphics[width=\textwidth]{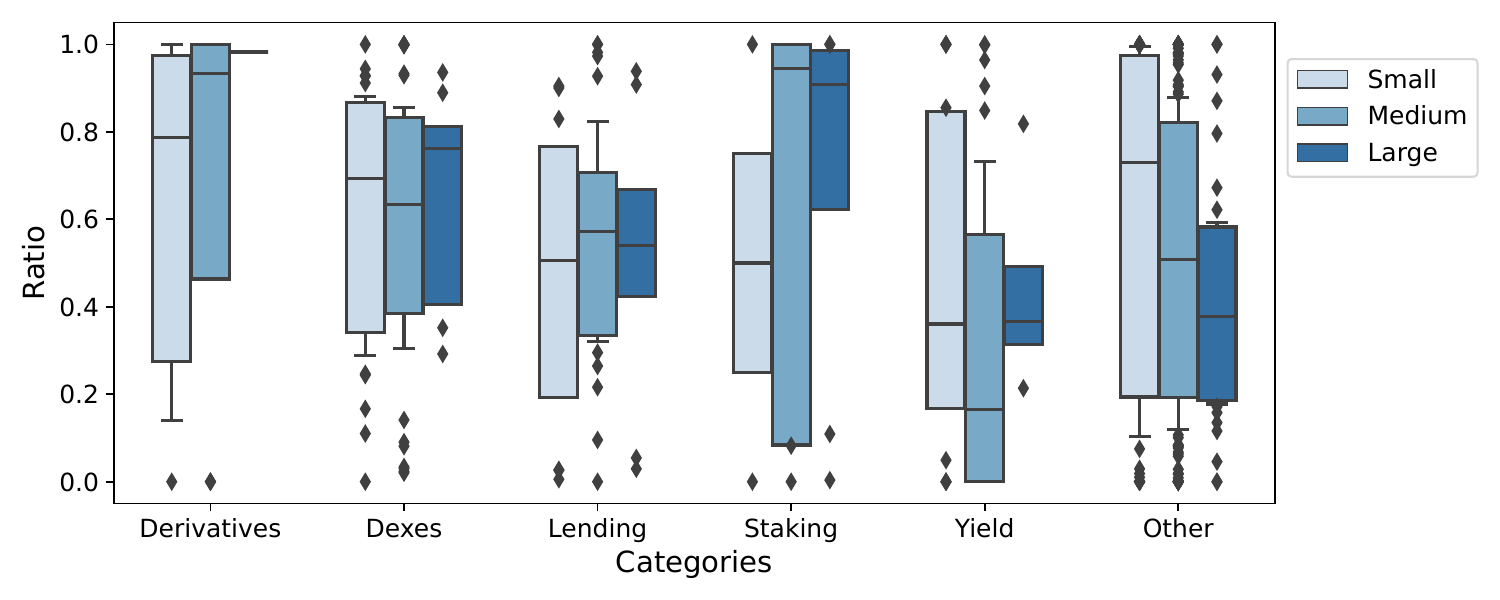}
		\hspace{0.5cm}
		\caption{}
		\label{fig:size}
	\end{subfigure}
	\vfill
	\begin{subfigure}[b]{\linewidth}
		\centering
		\includegraphics[width=\textwidth]{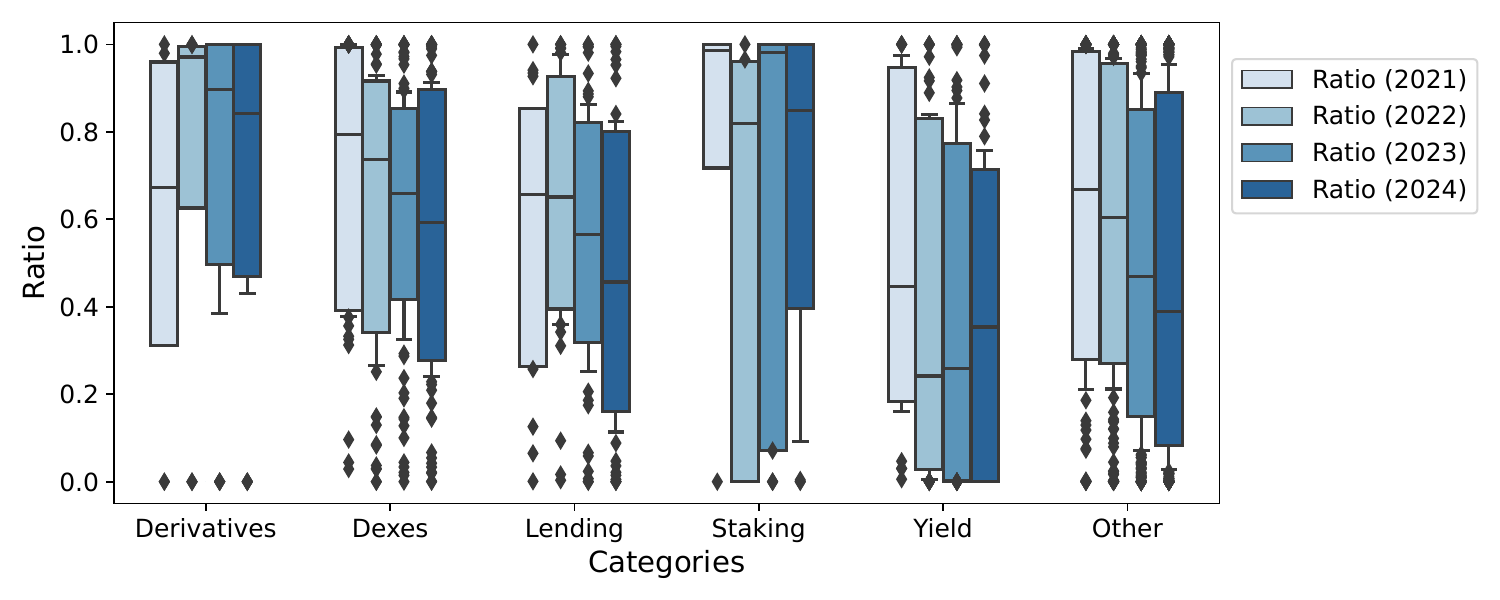}
		\caption{}
		\label{fig:time}
	\end{subfigure}
	\caption{\textbf{Ratio of Total Value Redeemable over Total Value Locked.} Protocols are grouped by category and further divided by size (\subref{fig:size}) and time (\subref{fig:time}). The y-axis indicates the ratio R between TVR and TVL. DEXs and Lending protocols rely more on non-redeemable tokens in 2024 w.r.t. earlier years.}
	\label{fig:boxplots}
\end{figure}

Figure \ref{fig:boxplots} shows two panels conveying additional information on the ratio R, with protocols further divided by size (upper panel) and time (lower panel). 
More precisely, on top we divide protocols in small size (N = 117), medium size (N = 232), and large size (N = 63), respectively when they have average TVL below $10^{6}$  USD, between $10^{6}$ USD and $10^{8}$  USD, and larger than $10^{8}$  USD. On the bottom, we split the R values for each protocol across time (2021, 2022, 2023, 2024). 
Interestingly, we do not find strong patterns when distinguishing protocols by size. Instead, in Panel~(\subref{fig:time}) we observe that for DEXs, Lending, and Derivatives protocols (excluding 2021), the median of R is steadily decreasing over time, indicating increased reliance on non-redeemable tokens.

\end{document}